\documentclass[a4paper,11pt]{article}
\pdfoutput=1 

\usepackage{jheppub} 

\usepackage[T1]{fontenc} 
\usepackage{blkarray}
\usepackage{relsize}

\usepackage{filecontents}

\def \trg {$\text{Tr\,G}_{\mathbb{I}}$}
\def \tg {$\text{Tr\,G}$}
\def \trga {$\text{Tr\,G}_{\alpha}$}

\title{\boldmath A Soft Theorem for the Tropical Grassmannian}

\author[a,b,c]{Diego Garc\'ia Sep\'ulveda}
\author[a,b,d]{and Alfredo Guevara}

\affiliation[a]{Perimeter Institute for Theoretical Physics, Waterloo, ON N2L 2Y5, Canada}
\affiliation[b]{Department of Physics and Astronomy, University of Waterloo, Waterloo, ON N2L 3G1, Canada}
\affiliation[c]{
ICTP South American Institute for Fundamental Research \\
Instituto de F\'isica Te\'orica, UNESP-Universidade Estadual Paulista \\
R. Dr. Bento T. Ferraz 271, Bl. II, S\~ao Paulo 01140-070, SP, Brazil}
\affiliation[d]{CECs Valdivia and Departamento de F\'isica, Universidad de Concepci\'on, Casilla 160-C, Concepci\'on, Chile}

\emailAdd{dgarcia@perimeterinstitute.ca}
\emailAdd{aguevara@perimeterinstitute.ca}

\abstract{We study the soft limit of a recently proposed generalization of the biadjoint scalar amplitudes $m^{(k)}_{n}$, which have been conjectured to have a relation to the tropical Grassmannian \tg($k$,$n$). Using the CHY formulation along with the Global Residue Theorem, we prove the soft factorization for $m^{(k)}_{n}$ amplitudes for arbitrary $k$ and $n$. We find that the soft factors are in direct correspondence to vertices of the associahedron $\mathcal{A}_{k-1}$, and hence take the form of $m^{(2)}_{n}$ amplitudes. This entails that all scattering amplitudes of the ordinary biadjoint scalar theory can be interpreted as an infinite family of soft factors. Additionally, Grassmannian duality reveals that generalized amplitudes $m^{(k)}_{n}$ with $k>2$ satisfy not only a soft theorem, but also a non-trivial ``hard'' theorem. We perform numerical checks of our theorems against previous results for \tg{(4,7)} and \tg{(5,8)}, thereby providing strong evidence of their relation with the CHY formulation.}

\begin{document} 
\maketitle

\flushbottom

\section{Introduction} \label{introduction}

In \cite{cachazo:2019ngv} Cachazo, Early, Mizera and one of the authors introduced a direct connection between scattering amplitudes and an intriguing structure known as the tropical Grassmannian \tg$(k,n)$ \cite{73806}. Such relation was originally found by generalizing the scattering equations to configuration spaces of $n$ points living in $\mathbb{CP}^{k-1}$, where $k=2$ leads to the original Cachazo-He-Yuan (CHY) formulation of Quantum Field Theory (QFT) amplitudes \cite{Cachazo:2013hca,Cachazo:2013iea,Cachazo:2014xea}. In this extension of the formalism one may consider the generalization of the simplest theory that is known to have a CHY form, namely, the biadjoint scalar theory. This cubic theory has been uncovered as the heart of many constructions recently studied in amplitudes, ranging from the KLT relations \cite{Kawai:1985xq,Broedel:2013tta,Cachazo:2013iea,Cachazo:2014xea,Mizera:2016jhj} to the kinematic associahedron \cite{Mizera:2017cqs,Arkani-Hamed:2017mur,Frost:2018djd,Salvatori:2018aha,Banerjee:2018tun,Raman:2019utu}.

The connection conjectured in \cite{cachazo:2019ngv} states that the amplitudes of the biadjoint theory for general $k$ are directly related to a subset of facets of the tropical Grassmannian \tg($k,n$). These objects are polyhedral fans whose rays correspond to kinematic invariants and whose facets can be mapped to Feynman diagrams, in the general $k$ sense. The most familiar case is the \tg(2,$n$) family which facets compute the standard biadjoint amplitudes $m^{(2)}_n$. On the other hand, it was observed by Arkani-Hamed, Bai, He and Yang (ABHY) \cite{Arkani-Hamed:2017mur} that such amplitudes correspond to the canonical form of the (kinematic) associahedron $\mathcal{A}^{\rm kin}_{n-3}$.\footnote{As the biadjoint theory arises as the low energy limit of disk integrals, the connection between biadjoint theory and the associahedron was introduced in \cite{Mizera:2017cqs} from a string theory perspective, see also \cite{Brown:2009qja}.} Such structure is indeed dual to the positive part of \tg(2,$n$) \cite{speyer2005tropical}.
For $k=3$ the generalized amplitudes $m^{(3)}_n$ were first obtained from \tg(3,6) in \cite{cachazo:2019ngv} and later from \tg(3,7) in \cite{Cachazo:2019apa}. On the other hand, a geometric interpretation was provided in \cite{Drummond:2019qjk}, which linked these amplitudes to certain cluster polytopes \cite{fomin2002cluster,fomin2003cluster,berenstein2005cluster,fomin2007cluster} dual to the tropical Grassmannians, therefore extending the associahedron interpretation of ABHY. As it turns out, for \tg(3,$n$) the $n=6,7,8$ amplitudes can be obtained from volumes of the $D_4, E_6$ and $E_8$ cluster polytopes respectively. In this way, the amplitudes given in \cite{cachazo:2019ngv,Cachazo:2019apa} were recovered, and a precise form of the biadjoint amplitude for \tg(3,8) was proposed. Note that due to the conjectured relation, we refer to the amplitudes of the generalized biadjoint scalar theory as tropical Grassmannian amplitudes interchangeably from now on.

Although the relation with the CHY formalism makes the tropical Grassmannian a promising candidate for constructing physical theories, it is still unclear how unitarity or locality emerge for these amplitudes. In this work we take a first step in this direction by  studying their soft limit \cite{Weinberg1965}. Focusing on $k=3$, we first argue that such operation can be defined in geometrical terms as an embedding of the ordered facets of \tg$(3,5)$ into \tg$(3,6)$. We then introduce the analytic definition of the limit in terms of kinematic invariants and study it from the CHY perspective. Indeed, one of the early successes of the $k=2$ CHY formulation was to make the well-known soft theorem of Weinberg completely straightforward to derive. This was important evidence supporting the conjectural CHY expression for gauge theory amplitudes, later proven in \cite{Dolan:2013isa}. 

We will show that tropical Grassmannian amplitudes indeed satisfy a soft theorem analogous to the $k=2$ case. Moreover, for $k>2$ we find a non-trivial ``hard'' theorem induced by the Grassmannian duality $m^{(k)}_{n}=m^{(n-k)}_{n} $. For any value of $k$ we will show that the soft (or ``hard'') factor is obtained from the positive facets of the smaller object \tg(2,$k+2$), or equivalently, from the vertices of the associahedron $\mathcal{A}^{\rm kin}_{k-1}$. In practice, this means that the soft factor is given explicitly by the ordinary cubic biadjoint amplitudes $m^{(2)}_{k+2}$ up to appropriate relabelings of its variables.

In order to outline our strategy let us briefly rederive the soft theorem for $k=2$ biadjoint scalar amplitudes, corresponding to \tg(2,$n$). These amplitudes are given by the integration over $n-3$ punctures on $\mathbb{CP}^1$ localized by  $n-3$ multidimensional contours $|E_i|=\epsilon $ \cite{Cachazo:2013iea, Cachazo:2013hca}, i.e.
\begin{equation}
m_n^{(2)}[\mathbb{I} | \mathbb{I}]=\int \prod_i^{n-3}{}' \frac{d\sigma_i}{E_i} \left(\frac{1}{\sigma_{12}\cdots \sigma_{n1}} \right)^2 \,,\quad \sigma_{ij}=\sigma_{i}-\sigma_j\,.
\end{equation}
Taking particle 1 to be soft, i.e. setting $s_{1i}=\tau \hat{s}_{1i}$ with $\tau \to 0$, the above formula behaves as
\begin{align}\label{eq:minisoft}
m_n^{(2)}[\mathbb{I}_n | \mathbb{I}_n]= & \frac{1}{\tau }\int \prod_{i\neq 1}^{n-4}{}' \frac{d\sigma_i}{E_i} \left(\frac{1}{\sigma_{23}\cdots \sigma_{n2}} \right)^2 \times \int \frac{d\sigma_1}{\hat{E}_1}\left(\frac{\sigma_{n2}}{\sigma_{n1}\sigma_{12}}\right)^2+\mathcal{O}(\tau^0) 
\end{align}
%
where $\hat{E}_1=\sum_{j\neq 1}\frac{\hat{s}_{1j}}{\sigma_{1j}}$. In the $\mathbb{CP}^1$ case, the integral over $\sigma_1$ corresponds to the soft factor and can be computed via a direct application of the residue theorem. Indeed, the contour $\hat{E}_1\to 0$ will be deformed to enclose the singularities of the integrand, where puncture $\sigma_1$ collides with either $\sigma_n$ or $\sigma_2$. In the $\mathbb{RP}^1$ projection, these lie at the boundary of the moduli associahedron $\mathcal{A}_{1}$ defined as the region $\{\sigma_1|\sigma_n<\sigma_1<\sigma_2\}$, see e.g. \cite{devadoss1999tessellations,Brown:2009qja}. Explicitly
\begin{align}
\label{basecase}
\tau S^{(2)}:=\int_{\hat{E}_1 \to 0} \frac{d\sigma_1}{\hat{E}_1}\left(\frac{\sigma_{n2}}{\sigma_{n1}\sigma_{12}}\right)^2 =& - \int_{\sigma_{n1}\to 0} \frac{d\sigma_1}{\hat{E}_1}\left(\frac{\sigma_{n2}}{\sigma_{n1}\sigma_{12}}\right)^2  - \int_{\sigma_{12}\to 0} \frac{d\sigma_1}{\hat{E}_1}\left(\frac{\sigma_{n2}}{\sigma_{n1}\sigma_{12}}\right)^2 \nonumber \\
=& \frac{1}{\hat{s}_{n1}}+ \frac{1}{\hat{s}_{12}}.
\end{align}
We see that the singularities in moduli space have been mapped to singularities in kinematic space, which are in fact vertices of the kinematic associahedron $\mathcal{A}^{\rm kin}_1$ of ABHY.
The soft factor corresponds to its canonical form, which is the standard four-point amplitude $m_4^{(2)}(\mathbb{I}_4|\mathbb{I}_4)=1/s_{41}+1/s_{12} $, once we perform the replacements $s_{41}\to \hat{s}_{n1}$ and $s_{12} \to \hat{s}_{12}$. From \eqref{eq:minisoft} we then obtain the standard soft theorem
\begin{equation}\label{eq:s2the}
    m_n^{(2)}[\mathbb{I}_n | \mathbb{I}_n] = S^{(2)}\times m_{n-1}^{(2)}[2\ldots n| 2\ldots n] + \mathcal{O}(\tau^0).
\end{equation}
In this work we will extend this result and show that for \tg($k,n$) the soft factor $S^{(k)}$ is given by $m_{k+2}^{(2)}$, the canonical form of $\mathcal{A}^{\rm kin}_{k-1}$, or alternatively by the facets of the positive \tg(2,$k+2$). 

To prove the above statements for $k>2$ is far more delicate and requires to implement multidimensional contour deformations using the Global Residue Theorem (GRT) \cite{griffiths2011principles}. We first prove directly the case $k=3$ and use it to gain some intuition for higher $k$: We find the precise mapping from singularities in the moduli associahedron $\mathcal{A}_{k-1}$ arising from the CHY integrand to boundaries of $\mathcal{A}^{\rm kin}_{k-1}$.

The definitive proof of the soft theorem for general $k$ is achieved in two steps. We first use the GRT to establish a recursion relation for the soft factor $S^{(k)}$ in terms of the lower soft factors $S^{(j)}$, with $j<k$. We then show that this recursion relation is satisfied by the biadjoint amplitude $m_{k+2}^{(2)}$ with appropriate relabelings.

We will check the soft factors against all explicit amplitudes available to us, including the ones given in \cite{Drummond:2019qjk}, up to $k=5, n=8$. Indeed, the soft theorem for \tg(5,8) can be also read as a hard theorem for \tg(3,8), and involves the non-trivial soft/hard factor presented in Appendix \ref{AppendixB}. This provides strong evidence on the conjectured relation between \tg($k,n$) and the CHY formalism.

This paper is organized in the following way: In section \ref{section2} we review the tropical Grassmannian \tg(3,6) and introduce a combinatorial description of its soft limit. In section \ref{section3} we review the CHY formalism for generalized biadjoint amplitudes and use it to provide a direct proof of the soft theorem for $k=3$. In section \ref{section4} we discuss the geometrical insight of this proof, thereby unveiling the relation between the soft factor and the associahedron for any $k$. Using this insight, and based on a recursion relation in $k$ given in Appendix \ref{appendixA}, in section \ref{section5} we provide the statement and outline the derivation of the soft theorem for general \tg$(k,n)$ amplitudes. We finish in section \ref{section6} with some discussions and directions for future research.

\section{The Soft Limit for \tg{(3,6)}} \label{section2}

In this section we first review the structure of the tropical Grassmannian \tg{(3,6)} \cite{73806}, and then proceed to construct a definition of the soft limit from a geometric perspective. After reviewing the tropical Grassmannian as a whole, we specialize to the facets associated to a given ordering. These have been revealed to have a direct connection to a generalization of biadjoint scalar theory first studied in \cite{cachazo:2019ngv}. We finish this section with a comment on how tropical Grassmannian duality allows us to consider a ``hard limit'' as the dual of the soft limit described in this section.

\subsection{Review of \tg{(3,6)}} \label{subsection2.1}

The tropical Grassmannian \tg{(3,6)} \cite{73806} is a polyhedral fan consisting of 65 vertices, 550 edges, 1395 triangles and 1035 tetrahedra. The set of vertices is obtained from solutions to tropical planes and consists of vectors (or rays) of the form
\begin{equation}\label{eq:vec}
W=(w_{123},w_{124},\ldots ,w_{456}) \in \mathbb{R}^{\binom{6}{3}}\,.
\end{equation}
The full set is given in \cite{73806}. In particular, the set includes the basis vectors 
\begin{equation}
\hat{e}_{ijk}=(0,\ldots ,1,\ldots,0) \,,
\end{equation}
with three unordered and unrepeated indices $i,j,k \in \{1,...,6 \}$. Considering the coordinate vector $S=(\mathbf{s}_{123},\mathbf{s}_{124},\ldots)$, we can define:
\begin{equation}
    S\cdot W=\sum_{i<j<k}\mathbf{s}_{ijk} w_{ijk}.
\end{equation}
We can now identify $S\cdot \hat{e}_{ijk} = \mathbf{s}_{ijk}$ as vertices. Since in the rest of this paper we work only with the latter objects $\mathbf{s}_{ijk}$, and therefore no confusion should arise, we use the name vertices for them from now on. Now, the vectors \eqref{eq:vec} posses a shift symmetry $w_{ijk} \to w_{ijk} + c_i + c_j+c_k$ inherited from their tropical planes. This implies the relation
\begin{equation}
\label{modout}
   0 = \sum_{\substack{j<k \\ j,k \neq i}} \mathbf{s}_{ijk}. 
\end{equation}
%
Besides $\mathbf{s}_{ijk}$, the remaining vertices are mapped to the following combinations:
\begin{equation}
\label{tintermsofs}
    \mathbf{t}_{abcd} = \mathbf{s}_{abc} + \mathbf{s}_{abd} + \mathbf{s}_{acd} + \mathbf{s}_{bcd},
\end{equation}
\begin{equation}
    R_{abcdef} = \mathbf{t}_{abcd} + \mathbf{s}_{cde} + \mathbf{s}_{cdf}.
\end{equation}

The following relations hold:
\begin{equation}
\label{identity}
    R_{12,34,56} + R_{34,12,56} = \mathbf{t}_{1234} + \mathbf{t}_{3456}+ \mathbf{t}_{5612},
\end{equation}
\begin{equation}
    R = R_{12,34,56} = R_{34,56,12} = R_{56,12,34}.
\end{equation}

There are $20$ vertices corresponding to $\mathbf{s}$ (forming a set $E$), 15 vertices corresponding to $\mathbf{t}$ (in a set $F$), and $30$ vertices corresponding to $R$ (in a set $G$). Starting from the vertices $v_i\in E \cup F \cup G$ one can then build the edges of the \tg({3,6}) as particular pairs $\{v_{i_h},v_{j_h}\}$. There are 550 such edges in \tg(3,6).

As shown in \cite{cachazo:2019ngv}, the generalized biadjoint amplitudes $m^{(3)}_6$ can be defined by facets (i.e. sets of edges) associated to given planar orderings. Let us describe first the full set of facets of \tg(3,6).

The facets come in six different classifications and 15 four-simplices. The six classifications correspond to tetrahedral facets\footnote{A tetrahedral facet is defined as a set of four vertices $\{ v_{1}, v_{2}, v_{3},v_{4}\} \subset E \cup F \cup G$, where each pair of vertices is an edge.}, given by all possible relabelings of the representatives shown in Table 1. The 15 four-simplices are given by permutations of labels of:
\begin{equation}
\label{four-15simplices}
    \{ \mathbf{t}_{1234}, \mathbf{t}_{3456}, \mathbf{t}_{5612}, R_{12,34,56},  R_{12,56,34}\},
\end{equation}
and we associate to them a representative contribution
\begin{equation}
\label{theta}
    \Theta = \frac{\mathbf{t}_{1234}+\mathbf{t}_{3456}+\mathbf{t}_{5612}}{\mathbf{t}_{1234}\mathbf{t}_{3456}\mathbf{t}_{5612} R \tilde{R}}, \ \ \ \ \  \tilde{R} = R_{(12) \leftrightarrow (34)}. 
\end{equation}
These contributions must be considered separately from the facets given by Table 1, and we must also consider them together with contributions $\tilde{\Theta}$, arising from \eqref{four-15simplices} under a cyclic shift of the labels. Notice that we can use (\ref{identity}) to rewrite (\ref{theta}) as a sum of either two or three terms,
\begin{align}
\label{eq:splittheta}
    \Theta & = \frac{1}{\mathbf{t}_{1234}\mathbf{t}_{3456}\mathbf{t}_{5612} R }+\frac{1}{\mathbf{t}_{1234}\mathbf{t}_{3456}\mathbf{t}_{5612} \tilde{R}} \nonumber \\
    & = \frac{1}{\mathbf{t}_{1234}\mathbf{t}_{3456} R \tilde{R}}+\frac{1}{\mathbf{t}_{1234}\mathbf{t}_{5612} R \tilde{R}}+\frac{1}{\mathbf{t}_{3456}\mathbf{t}_{5612} R \tilde{R}}.
\end{align}

As explained in \cite{cachazo:2019ngv} (see also \cite{Drummond:2019qjk}), these correspond to two different ways to split a bipyramid in terms of tetrahedra. Eq. (\ref{theta}) makes manifest the fact that the fundamental object is the bipyramid, and not the tetrahedra in which it can be split (and that are not facets of \tg{(3,6)}).


With this information one can give an algorithm to construct $m^{(3)}_{6}[\alpha|\beta]$ solely from \tg{(3,6)}, where $\alpha$ and $\beta$ are two permutations of the cyclic ordering $\mathbb{I}_6=(123456)$. One takes the subset of all vertices in $E \cup F \cup G$ consistent with a given permutation $\alpha$.  This defines an ordered part of the tropical Grassmannian which we refer to as \trga{(3,6)}. The corresponding facets are the subset of facets of \tg{(3,6)} made with the vertices of \trga{(3,6)}. Following \cite{cachazo:2019ngv} we denote their contribution by $J(\alpha)$, where each facet is represented by the inverse power of its vertices. For instance

\begin{equation}
J(\mathbb{I}_6)=\{ \frac{1}{\mathbf{s}_{123}\mathbf{s}_{345} \mathbf{s}_{561} \mathbf{s}_{246}}, \frac{1}{\mathbf{s}_{123}\mathbf{s}_{456}\mathbf{t}_{1234}\mathbf{t}_{3456}},\frac{1}{\mathbf{s}_{345} \mathbf{t}_{1256}\mathbf{t}_{3456} R_{12,34,56}}\,,\ldots ,\Theta_{123456},\tilde{\Theta}_{123456}\}
\end{equation}
has 48 elements. Then 
\begin{equation}\label{eq:compt}
    m^{(3)}_{6}[\alpha|\beta] = \sum_{\Upsilon \in J(\alpha) \bigcap  J(\beta)} \Upsilon \,.
\end{equation}
We will be mostly interested in the full amplitude of \trg(3,6), given by the set of all facets consistent with the canonical ordering. This is given by
\begin{equation}\label{eq:defm}
    m^{(3)}_{6}[\mathbb{I}_6|\mathbb{I}_6] = \sum_{\Upsilon \in J(\mathbb{I}_6)} \Upsilon\,.
\end{equation}
Let us exemplify the procedure just described using the orderings $\alpha = \mathbb{I}_{6}$, $\beta = 125643$. It is straightforward to see that these orderings are consistent with the vertices (\ref{four-15simplices}) defining a four-simplex, so we must consider a $\Theta$ contribution (\ref{theta}), i.e. $\Theta_{123456}=\Theta_{125634}$. One can check that there are only three more facets that are consistent with the orderings $\alpha$ and $\beta$. One of these facets belong to the EEFF1 category, and the two remaining facets correspond to the EFFG category. The algorithm gives (up to an overall sign determined in \cite{cachazo:2019ngv}):

\begin{equation}
\label{exampleamplitude}
    m^{(3)}_{6}[\mathbb{I}_{6}|125643] = \Theta + \frac{1}{\mathbf{t}_{1234}\mathbf{t}_{3456}} \bigg( \frac{1}{\mathbf{s}_{456} \tilde{R} } + \frac{1}{\mathbf{s}_{123} R } + \frac{1}{\mathbf{s}_{123}\mathbf{s}_{456}}\bigg).
\end{equation}

\begin{table}[]
\begin{center}
    \begin{tabular}{ | l | l | l | p{5cm} |}
    \hline
    EEEE &  \{$\mathbf{s}_{123}$, $\mathbf{s}_{345}$, $\mathbf{s}_{561}$ ,$\mathbf{s}_{246}\}$ \\ \hline
    EEFF1 & \{$\mathbf{s}_{123}$, $\mathbf{s}_{456}$, $\mathbf{t}_{1234}$, $\mathbf{t}_{3456}\}$ \\ \hline
    EEFF2 & \{$\mathbf{s}_{123}$, $\mathbf{s}_{345}$, $\mathbf{t}_{3456}$, $\mathbf{t}_{1236}\}$  \\ \hline
    EFFG & \{$\mathbf{s}_{345}$, $\mathbf{t}_{1256}$, $\mathbf{t}_{3456}$, $R_{12,34,56}$\}  \\ \hline
    EEEG & \{$\mathbf{s}_{123}$, $\mathbf{s}_{561}$, $\mathbf{s}_{345}$, $R_{45,23,61}$\}  \\ \hline
    EEFG & \{$\mathbf{s}_{123}$, $\mathbf{t}_{3456}$, $\mathbf{s}_{345}$, $R_{12,34,56}$\}  \\ \hline
    \hline
    \end{tabular}
\caption{Representatives of facets of the tropical Grassmannian Tr\,G(3,6)}
\end{center}
\end{table}

\subsection{Embedding \trg{(3,5)} into \trg{(3,6)}} \label{subsection2.2}

Although we have provided the geometric structure of the tropical Grassmannian \tg(3,6), it is not yet clear what is the interpretation for a single label $i=1,\ldots,6$. Since for $k=2$ these labels correspond to the particles involved in a scattering process, let us attempt to extend such notion for higher values of $k$. With this in mind, and with the benefit of hindsight, let us consider the following set of vertices in the canonically ordered part of the tropical Grassmannian, \trg(3,6): 
\begin{equation}
\label{softvertices}
\big\{\mathbf{s}_{123},\mathbf{s}_{156},\mathbf{s}_{126},  \mathbf{t}_{2345},\mathbf{t}_{3456}\big\} \,.
\end{equation}
%
%
Consider now the edges inside \trg(3,6) constructed from this set of vertices. One finds that there are five such edges, and they turn out to form a smaller object, \trg(3,5)! The facets of this object are constructed from two vertices \cite{73806}, and they are characterized by the following contributions:
\begin{equation}
\label{edges}
    S_{1,23456}= \bigg\{\frac{1}{\mathbf{s}_{123}\mathbf{s}_{156}},  \frac{1}{\mathbf{t}_{2345}\mathbf{s}_{156}} , \frac{1}{\mathbf{t}_{2345}\mathbf{s}_{126}}  ,  \frac{1}{\mathbf{t}_{3456} \mathbf{s}_{123}} , \frac{1}{\mathbf{t}_{3456}\mathbf{s}_{126}}\bigg\}.
\end{equation}
For latter convenience let us also define 
\begin{equation}
\label{eq:defs3}
    S^{(3)}_{1,23456}=  \sum_{\Upsilon \in S_{1,23456}} \Upsilon = \frac{1}{\mathbf{s}_{123}\mathbf{s}_{156}} + \frac{1}{\mathbf{t}_{2345}} \bigg(\frac{1}{\mathbf{s}_{156}} + \frac{1}{\mathbf{s}_{126}} \bigg) + \frac{1}{\mathbf{t}_{3456}} \bigg(\frac{1}{\mathbf{s}_{123}} + \frac{1}{\mathbf{s}_{126}}\bigg).
\end{equation}
Since $S_{1,23456}$ is identified with a subset of edges of \trg(3,6), the corresponding edges will naturally appear in the facets of \trg(3,6). In fact, each facet turns out to have at most one edge of the set $ S_{1,23456}$. A way to show this explicitly is to write down the amplitude \eqref{eq:defm}, which contains the following terms:
\begin{align} \label{someterms}
       m^{(3)}_{6}[\mathbb{I}_{6}|\mathbb{I}_{6}]
       &=\frac{1}{\mathbf{s}_{123}\mathbf{s}_{561}}\left[\frac{1}{R_{45,23,61}\mathbf{s}_{345}}+\frac{\frac{1}{\mathbf{s}_{345}}+\frac{1}{\mathbf{t}_{1234}}}{R_{12,34,56}}+\frac{\frac{1}{R_{45,23,61}}+\frac{1}{\mathbf{t}_{1234}}}{\mathbf{t}_{4561}}\right]\nonumber\\
       & +\frac{1}{\mathbf{t}_{2345}\mathbf{s}_{561}}\left[\frac{1}{R_{45,23,61}\mathbf{s}_{345}}+\frac{\frac{1}{R_{45,23,61}}+\frac{1}{\mathbf{s}_{ 234}}}{\mathbf{t}_{4561}}+\frac{\frac{1}{\mathbf{s}_{234}}+\frac{1}{\mathbf{s}_{345}}}{\mathbf{t}_{5612}}\right]\nonumber \\
       & +\frac{1}{\mathbf{t}_{2345}  \mathbf{s}_{612}} \left[\frac{1}{R_{23,45,61} s_{234}}+\frac{\frac{1}{\mathbf{s}_{234}}+\frac{1}{\mathbf{s}_{345}}}{\mathbf{t}_{5612}}+\frac{\frac{1}{R_{23,45,61}}+\frac{1}{\mathbf{s}_{345}}}{\mathbf{t}_{6123}} \right]\nonumber \\
       & + \frac{1}{\mathbf{t}_{3456}\mathbf{s}_{123}} \left[ \frac{1}{R_{12,34,56}
       \mathbf{s}_{345}}+\frac{\frac{1}{R_{12,34,56}}+\frac{1}{\mathbf{s}_{456}}}{\mathbf{t}_{1234}}+\frac{\frac{1}{\mathbf{s}_{345}}+\frac{1}{\mathbf{s}_{456}}}{\mathbf{t}_{6123}} \right]\nonumber \\
       & + \frac{1}{\mathbf{t}_{3456}\mathbf{s}_{612}}  \left[\frac{1}{R_{34,12,56}
       \mathbf{s}_{456}}+\frac{\frac{1}{R_{34,12,56}}+\frac{1}{\mathbf{s}_{345}}}{\mathbf{t}_{5612}}+\frac{\frac{1}{\mathbf{s}_{345}}+\frac{1}{\mathbf{s}_{456}}}{\mathbf{t}_{6123}}\right] +\ldots 
\end{align}
We find that there are five groups of five terms, each group containing a given edge in (\ref{edges}). Quite non-trivially, it turns out that these also contain the structure of \eqref{eq:defs3}. This means each group has the structure of \trg(3,5) but it is constructed from a different set of vertices. In fact, these groups are nothing but copies of the smaller amplitude $m^{(3)}_5$! To see this we introduce the following limit, which we call the soft limit associated to the label 1:
\begin{equation}
\label{softlimit}
    \mathbf{s}_{1ab} = \tau \hat{\mathbf{s}}_{1ab}, \ \ \ \tau \rightarrow 0, \ \ \ \hat{\mathbf{s}}_{1ab} \,\,\,\text{fixed}.
\end{equation}
From all the vertices in \trg(3,6), only the set of vertices (\ref{softvertices}) vanish at $\tau=0$, which is the reason why we considered this set to begin with. We refer to the vertices in this set as the ``soft vertices'' associated to the label 1 and to the canonical ordering. The edges inside $S_{1,23456}$ are then  called ``soft edges''. The remaining ``hard'' vertices are deformed and possibly degenerate into each other. For example, $\mathbf{t}_{1234}\to \mathbf{s}_{234}$, $R_{23,45,61}\to \mathbf{s}_{456}$, etc...

We leave for future work the interpretation of this deformation in terms of the tropical hyperplanes. For now, we simply note that in the limit \eqref{softlimit} the square brackets in \eqref{someterms} collapse to the same quantity. Note further that the terms in  $m^{(3)}_{6}[\mathbb{I}_{6}|\mathbb{I}_{6}]$ not containing a soft edge will be subleading as $\tau \to 0$. Hence, the amplitude exhibits the remarkable factorization

\begin{align}
    m^{(3)}_{6}[\mathbb{I}_{6}|\mathbb{I}_{6}] =&\,\,
    \frac{1}{\tau^{2}}\bigg[ \frac{1}{\hat{\mathbf{s}}_{123}\hat{\mathbf{s}}_{156}} + \frac{1}{\hat{\mathbf{t}}_{2345}} \bigg(\frac{1}{\hat{\mathbf{s}}_{156}} + \frac{1}{\hat{\mathbf{s}}_{126}} \bigg) + \frac{1}{\hat{\mathbf{t}}_{3456}} \bigg(\frac{1}{\hat{\mathbf{s}}_{123}} + \frac{1}{\hat{\mathbf{s}}_{126}}\bigg) \bigg]\nonumber \\
   & \times  \bigg[ \frac{1}{\mathbf{s}_{236}\mathbf{s}_{345}} +
    \frac{1}{\mathbf{s}_{256}} \bigg( \frac{1}{\mathbf{s}_{345}} + \frac{1}{\mathbf{s}_{234}} \bigg) + \frac{1}{\mathbf{s}_{456}} \bigg( \frac{1}{\mathbf{s}_{236}} + \frac{1}{\mathbf{s}_{234}}  \bigg)  \bigg] + \mathcal{O}\Big( \frac{1}{\tau} \Big)\nonumber \\
    = & \,\, S^{(3)}_{1,23456} \times \,m^{(3)}_{5}[23456|23456 ] + \mathcal{O}\Big( \frac{1}{\tau} \Big) \,, \label{eq:s36theo}
\end{align}
where we used that the second factor also corresponds to a $k=3$ amplitude \cite{cachazo:2019ngv}, independent of label 1. This is a generalization of the $k=2$ soft theorem \eqref{eq:s2the} to the case of \trg(3,6). Considering this generalization, we refer to  $S^{(3)}_{1,23456}$ in (\ref{eq:defs3}) as a soft factor. We are going to prove that this factorization holds for all $m^{(3)}_n$ amplitudes in section \ref{section3}. We extend the theorem to arbitrary values of $k$, and prove it in section \ref{section5} and appendix \ref{appendixA} respectively. Note that in contrast to the $k=2$ case, the leading divergence here behaves as $\sim \frac{1}{\tau^2}$. 

\begin{figure}
  \centering
    \includegraphics[width=0.9\textwidth]{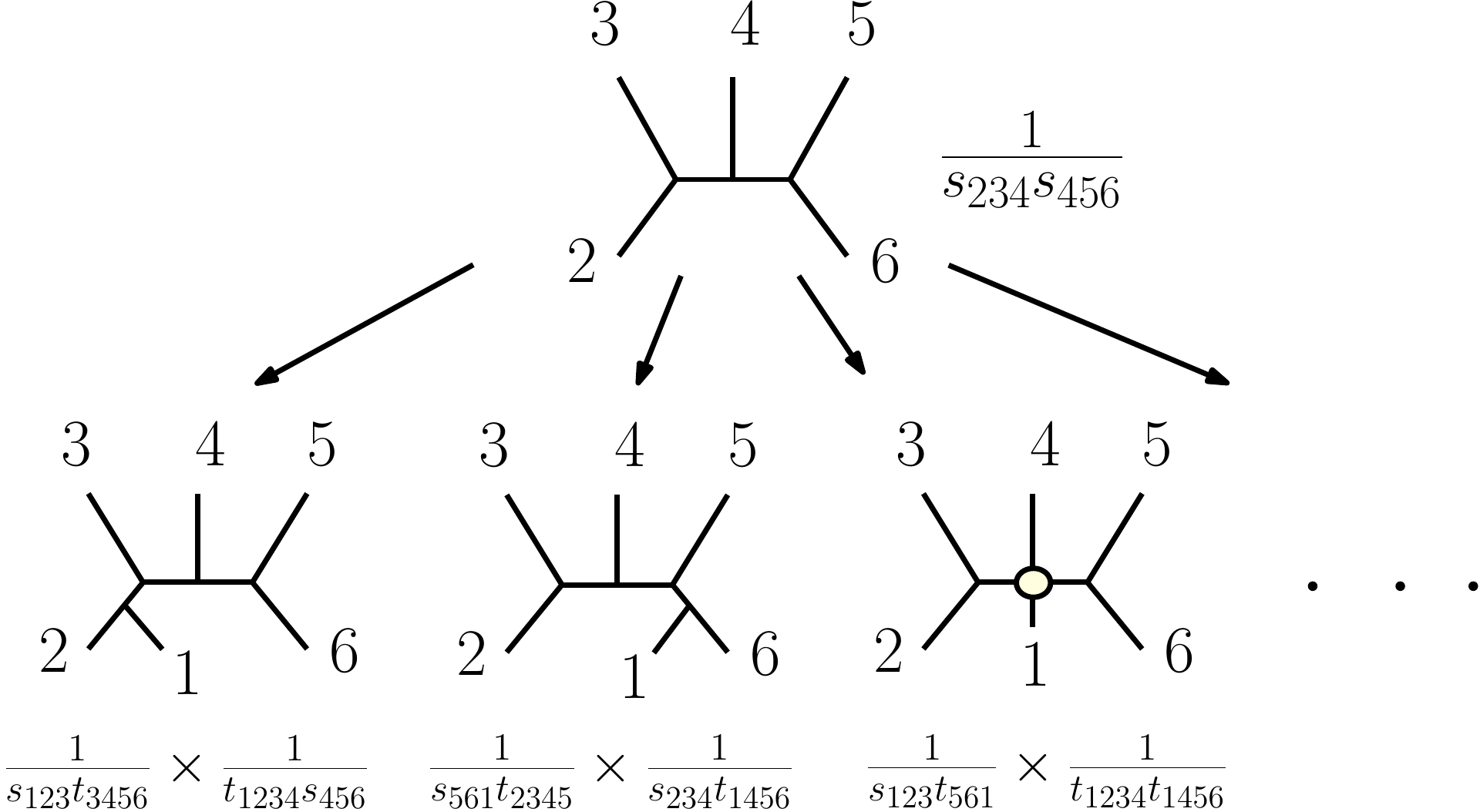}
      \caption{Diagrammatic representation of the embedding of a facet of $m^{(3)}_5$ into five facets of $m^{(3)}_6$, one for each soft edge (symmetries of the diagram are symmetries of the term). This can be thought as attaching label $1$ at different positions. Note that for $k=2$ (i.e. biadjoint theory) only the first two diagrams would appear in the soft limit.}
      \label{fig:fg1}
\end{figure}

In terms of the facets, the soft theorem can be interpreted as an embedding 
\begin{equation}\label{eq:emb}
    J(\mathbb{I}_6) \supset  J_{\rm soft}(\mathbb{I}_6) \sim   S_{1,23456} \times  J(23456),
\end{equation}
where $J_{\rm soft}(\mathbb{I}_6)$ denotes the facets in $J(\mathbb{I}_6)$ containing a soft edge and $\times$ is here the cartesian product. This means each facet in $J(23456)$ can be mapped to a facet of $J(\mathbb{I}_6)$ in five different ways, one for each soft edge, see Figure \ref{fig:fg1}. Note that the two factors in \eqref{eq:emb} correspond to two instances of \trg(3,5).

Note that the formula \eqref{eq:s36theo} hints that the terms can be grouped as in \eqref{someterms}. This is because the soft factor $ S^{(3)}_{1,23456} $ is invariant under the soft limit, whereas for each soft edge the amplitude $m^{(3)}_{5}[23456|23456]$ will undergo a particular deformation as we depart from the $\tau \rightarrow 0$ limit. Additionally, the fact  that two or more soft edges cannot appear together in a facet follows here from the leading order of $m^{(3)}_6$ being $\tau^{-2}$, which is true for all $m^{(3)}_n$ amplitudes. On the other hand, one should be careful when analyzing facets with non-trivial numerators such as $\Theta$ in \eqref{theta} (see \cite{Cachazo:2019apa} for $n=7$ examples), which here turned out to not contribute to the soft limit as they do not contain soft edges.

The factorization \eqref{eq:emb} is not a property of $m^{(3)}_{6}[\mathbb{I}_{6}|\mathbb{I}_{6}]$ only. Rather, it is a property shown by any $m^{(3)}_{6}[\alpha|\beta]$, with $\alpha$ and $\beta$ any orderings. The factorization for arbitrary orderings is given by the following compatibility rule. For each ordering $(1\gamma)$, there is a set
\begin{equation}
\label{5in6}
    J_{\rm soft}(1\gamma) \sim S_{1,\gamma} \times J(\gamma).
\end{equation}
The left hand side of (\ref{5in6}) can be defined as the set of all facets in \trg(3,6) of order $\tau^{-2}$, and can be thought as evaluated at leading order. According to the definition \eqref{eq:compt} the order $\tau^{-2}$ facets of $m^{(3)}_{6}[\alpha|\beta]$ will be given by
\begin{align}
    J_{\rm soft}(1\alpha) \cap   J_{\rm soft}(1\beta)   \sim & \Big( S_{1,\alpha} \times J(\alpha) \Big) \cap \Big( S_{1,\beta} \times J(\beta) \Big) \nonumber \\
    =&  \Big( S_{1,\alpha} \cap S_{1,\beta} \Big) \times \Big( J(\alpha)  \cap  J(\beta) \Big).
\end{align}

This means that the leading order of $m^{(3)}_{6}[1\alpha|1\beta]$ also satisfies a soft theorem \eqref{eq:s36theo} with the soft factor
\begin{equation}
    S^{(3)}_{1\alpha|1\beta} =  \sum_{\Upsilon \in S_{1\alpha}\cap S_{1\beta} } \Upsilon \,,
\end{equation}
i.e. only considering soft edges compatible with both orderings. This rule can be trivially extended to all $m^{(k)}_n$ amplitudes, once we provide the corresponding soft theorems.

Let us provide a short example, and apply the rule to find the soft factorization associated to the amplitude $m^{(3)}_{6}[\mathbb{I}_{6}|125643]$. It is straightforward to see from (\ref{eq:defs3}) that there is only one soft edge consistent with the $125643$ ordering, namely, the $(\mathbf{t}_{3456} \mathbf{s}_{123})^{-1}$ term. On the other hand, we have \cite{cachazo:2019ngv}
\begin{equation}
     m^{(3)}_{5}[\mathbb{I}_{5}|25643] =  \frac{1}{\mathbf{s}_{234}}
    \bigg[\frac{1}{\mathbf{s}_{256}} + \frac{1}{\mathbf{s}_{456}}\bigg].
\end{equation}
We can check from (\ref{exampleamplitude}) that the $n=6$ amplitude indeed satisfies the claimed soft theorem:
\begin{equation}
\label{examplefactorization}
    m^{(3)}_{6}[\mathbb{I}_{6}|125643] \longrightarrow  \frac{1}{\tau^2}\bigg[ \frac{1}{\mathbf{\hat{t}}_{3456} \mathbf{\hat{s}}_{123}} \bigg] \times  \frac{1}{\mathbf{s}_{234}}
    \bigg[\frac{1}{\mathbf{s}_{256}} + \frac{1}{\mathbf{s}_{456}}\bigg] \,.
\end{equation}




Before closing this section, let us study a different soft factorization which we will further elaborate on the next sections. Consider the deformation
\begin{equation}
\label{hardlimit2}
    \mathbf{s}_{abc} = \tau \hat{\mathbf{s}}_{abc}, \ \ \ \tau \rightarrow 0, \ \ \  \hat{\mathbf{s}}_{abc} \ \mathrm{fixed}, \ \ \ a,b,c \neq 1,
\end{equation}
which we can interpret as a ``hard limit'' associated to the label 1, since now all vertices corresponding to this label remain finite while other vertices tend to zero. We find that in this limit the amplitude also factorizes, leading to 
\begin{equation*}
       m^{(3)}_{6}[\mathbb{I}_{6}|\mathbb{I}_{6}]  =\,\,
    \frac{1}{\tau^{2}}\bigg[ \frac{1}{\hat{\mathbf{s}}_{456}\hat{\mathbf{s}}_{234}} + \frac{1}{\hat{\mathbf{t}}_{2345}} \bigg(\frac{1}{\hat{\mathbf{s}}_{234}} + \frac{1}{\hat{\mathbf{s}}_{345}} \bigg) + \frac{1}{\hat{\mathbf{t}}_{3456}} \bigg(\frac{1}{\hat{\mathbf{s}}_{456}} + \frac{1}{\hat{\mathbf{s}}_{345}}\bigg) \bigg]\nonumber 
\end{equation*}
\begin{equation}
    \times  \bigg[ \frac{1}{\mathbf{s}_{145}\mathbf{s}_{126}} +
    \frac{1}{\mathbf{s}_{134}} \bigg( \frac{1}{\mathbf{s}_{126}} + \frac{1}{\mathbf{s}_{156}} \bigg) + \frac{1}{\mathbf{s}_{123}} \bigg( \frac{1}{\mathbf{s}_{145}} + \frac{1}{\mathbf{s}_{156}}  \bigg)  \bigg] + \mathcal{O}\Big( \frac{1}{\tau} \Big).
\end{equation}
We call this a ``hard'' theorem for \trg(3,6). Even though it looks unrelated at first to the soft theorem, it corresponds to a different soft limit: The limit of $m^{(3)}_{6}$ when dualized in the Grassmannian sense. We will come back to this remarkable property in the next section. For now we just point out that the fact that we find again two factors with the structure of \trg(2,5) is a consequence of the self-duality of \tg(3,6).

\section{Extended CHY and a Soft Theorem for all \trg(3,$n$)} \label{section3}

As we have already outlined the soft theorem for \trg(3,6) and its consequences, we shall now focus in proving the statement \eqref{eq:s36theo} for general \trg$(k,n)$. We start from the simplest case $k=3$. Since our work extensively employs the generalization of the scattering equations to $\mathbb{C}\mathbb{P}^{k-1}$ introduced in \cite{cachazo:2019ngv}, we first review such extension together with the corresponding generalized biadjoint amplitudes. We proceed then to construct the soft limit in such formalism. We close this section with a direct proof of the soft theorem for $k=3$ amplitudes via the GRT, thereby recovering \eqref{eq:s36theo} and extending it to all multiplicity. This provides non-trivial evidence for the conjectured relation between CHY and \trg(3,$n$), complementary to that of \cite{cachazo:2019ngv,Cachazo:2019apa}.

\subsection{Scattering Equations over $\mathbb{CP}^{k-1}$ and Grassmannian Duality} \label{subsection3.1}

In the extended CHY formalism, we define objects $\mathbf{s}_{a_{1}a_{2}...a_{k}}$ as completely symmetric tensors of rank $k$. The corresponding scattering equations can be obtained as the conditions for the extrema of the following potential function \cite{cachazo:2019ngv,Cachazo:2016ror}:

\begin{equation}
\label{CHYpotential}
    \mathcal{S}_{k} = \sum_{1 \leq a_{1} < a_{2} < ...<a_{k} \leq n} \mathbf{s}_{a_{1}a_{2}...a_{k}} \log{(a_{1} a_{2}...a_{k})},
\end{equation}
where $(a_{1} a_{2}...a_{k})$  are $\mathrm{SL(k,\mathbb{C})}$ invariant determinants of the homogeneous coordinates of the points $a_{i}$. Inhomogeneous coordinates are defined by rescaling the first homogeneous coordinate to be one, so inhomogeneous coordinates can be written as $(1,x^{1},\ldots,x^{k-1})$.

In order for the potential $\mathcal{S}_{k}$ to be well defined, the following condition must be satisfied:
\begin{equation}
\label{massless}
    \mathbf{s}_{a_{1}a_{2}...bb} = 0.
\end{equation}
The projective invariance of $\mathcal{S}_{k}$ implies
\begin{equation}
\label{momcons}
    \sum_{\substack{a_{2},a_{3},...,a_{k} = 1 \\ a_{i} \neq a_{j}}}^{n} \mathbf{s}_{a_{1}a_{2}...a_{k}} = 0, \ \ \ \ \ \ \forall a_{1}.
\end{equation}
If we apply this procedure for $k=2$ we find that ($\ref{massless}$) corresponds to the condition for a particle to be massless, and (\ref{momcons}) corresponds to the momentum conservation condition for kinematic invariants in QFT. Due to the fact that the abstract objects $\mathbf{s}_{a_{1}a_{2}...a_{k}}$ satisfy these properties, we identify them with kinematic invariants and we refer to their indices as particle labels from now on. This observation provides us with a novel conceptual understanding of the massless property and the principle of momentum conservation: they are direct consequences of a well-defined, projective invariant potential $\mathcal{S}_{k}$.

As we have made obvious through our notation, we have drawn upon the conjecture arising from the work \cite{cachazo:2019ngv} to call the kinematic invariants $\mathbf{s}_{a_{1}a_{2}...a_{k}}$ in the same way as the coordinate functions of \tg{($k$,$n$)} \cite{73806}. Thus the vertices of the tropical Grassmannian will in general correspond to specific combinations of coordinates as we saw for \tg(3,6). Along these lines, we notice that the momentum conservation condition (\ref{momcons}) corresponds to the  modding-out condition (\ref{modout}) that defines the tropical Grassmannian.

The configuration space over which the scattering equations are defined is denoted as $X(k,n)$, and it can be written as the Grassmannian G($k$,$n$), modded out by a $n$-torus action
\begin{equation}
    X(k,n) = G(k,n)/(\mathbb{C}^{*})^{n}.
\end{equation}

With all these ingredients, we can define the generalized biadjoint scalar amplitudes:
\begin{equation}\label{CHYm}
     m_{n}^{(k)}[\alpha | \beta] = \mathlarger{\int} \Bigg( \frac{1}{\text{Vol}[\mathrm{SL}(k,\mathbb{C})]} \prod_{a=1}^{n} \prod_{i=1}^{k-1} dx^{i}_{a} \Bigg) \ \prod_{a=1}^{n} \sideset{}{'}\prod_{i=1}^{k-1} \delta \bigg( \frac{\partial \mathcal{S}_{k}}{\partial x^{i}_{a}} \bigg) \times  \ \mathrm{PT}^{(k)}_{n}[\alpha] \  \mathrm{PT}^{(k)}_{n}[\beta],
\end{equation}
where $x^{i}_{a}$ stands for the $i$-th inhomogeneous coordinate of the $a$-th particle. As each particle lives in $\mathbb{CP}^{k-1}$ the integral must be $\mathrm{SL}(k,\mathbb{C})$ invariant. We can then fix the location of $k+1$ punctures to generic positions (i.e. with non-vanishing brackets), and similarly remove $k+1$ delta functions with a corresponding jacobian. The generalized Parke-Taylor ``half-integrand'' is given by:
\begin{equation}
     \mathrm{PT}^{(k)}_{n}[1,2,...,n]=  \frac{1}{|12\cdot \cdot \cdot k||23 \cdot \cdot \cdot (k+1)| ...  |n \ 1 \cdot \cdot \cdot  (k-1)|}.
\end{equation}

Now, it was observed in \cite{cachazo:2019ngv} that the potential \eqref{CHYpotential} inherits Grassmannian duality $G(k,n)\sim G(n-k,n)$. In particular, for a canonical chart of $X(k,n), \, X(n-k,n)$, we can identify the coordinates 
\begin{align}\label{duality}
    s_{a_1\ldots a_{n-k}} =& \,s_{b_1\ldots b_k}, \nonumber \\
    (a_1\ldots a_{n-k}) =& \,(b_1\ldots b_k),
\end{align}
for complement labels, i.e. $\{b_i\}\cup \{a_i\}=\{1,\ldots,n\}$. In this case we can write
\begin{equation}
\mathrm{PT}^{(k)}_{n}[1,2,...,n] = \mathrm{PT}^{(n-k)}_{n}[1,2,...,n].
\end{equation}
Finally, as the CHY formula must be independent of the $\text{SL}(k,\mathbb{C})$ fixing, it must be that
\begin{equation}
m^{(k)}_n [\alpha | \beta] = m^{(n-k)}_n [\alpha | \beta].
\end{equation}
In particular $m^{(2)}_{k+2} = m^{(k)}_{k+2}$, which will prove to be a key insight in constructing the soft factor.

\subsection{Soft Limit Setup} \label{subsection{3.2}}


Before presenting and proving the general soft theorem, we set the grounds by defining the soft limit in terms of the CHY formula. Throughout this subsection, and in the rest of the paper, we take particle 1 as the soft particle.

In the ordinary $k=2$ case, we define the soft limit by taking the low energy-momentum limit of a particle particle; that is, we define the soft momentum by $p_1^\mu = \tau \hat{p}_1^\mu$, (together with the corresponding deformations from momentum conservation), and then take $\tau \to 0$. For a scalar theory this is equivalent to set $\mathbf{s}_{1a} =  \tau \hat{\mathbf{s}}_{1a}$. For arbitrary $k$ one defines the soft limit as the natural generalization of the latter procedure: We take $\mathbf{s}_{1 a_{2}...a_{k}} = \tau \hat{\mathbf{s}}_{1 a_{2}...a_{k}}$, with $\tau \rightarrow 0$ and $\hat{\mathbf{s}}_{1 a_{2}...a_{k}}$ fixed, as done in section \ref{section2} for $k=3$. 

Before taking the soft limit in the CHY amplitude, let us review what happens to the scattering equations and momentum conservation in the soft limit. This has already been considered in \cite{cachazo:2019ngv}. To leading order in the soft limit, the momentum conservation condition for the soft particle decouples from the momentum conservation condition for the other particles, and we obtain the following two separate relations:
\begin{equation}
    \sum_{\substack{a_{2},a_{3},...,a_{k} = 2 \\ a_{i} \neq a_{j}}}^{n} \mathbf{s}_{a_{1}a_{2}...a_{k}} = 0, \ \ \ \ \ \  a_{1} \neq 1.
\end{equation}
\begin{equation}\label{softcon}
    \sum_{\substack{a_{2},a_{3},...,a_{k} = 2 \\ a_{i} \neq a_{j}}}^{n} \mathbf{s}_{1 a_{2}...a_{k}} = 0.
\end{equation}
Observe that to obtain these we take the variables $\mathbf{s}_{1 a_{2}...a_{k}}$ as independent (up to the condition \eqref{softcon}), which will be relevant when defining the hard limit.
In a similar way, the scattering equations for the soft particle separate to leading order from the scattering equations for the other particles:

\begin{equation}
\label{softscatteringeq}
    E^{x^{i}}_{1} \rightarrow  \tau \hat{E}^{x^{i}}_{1}, \ \ \ \ \ \ \hat{E}^{x^{i}}_{1} = \sum_{\substack{a_{2},a_{3},...,a_{k} = 2 \\ a_{i} \neq a_{j}}}^{n} \frac{\hat{\mathbf{s}}_{1 a_{2}...a_{k}}}{|1a_{2}...a_{k}|} \frac{\partial}{\partial x^{i}} |1a_{2}...a_{k}| = 0,
\end{equation}

\begin{equation}
    E^{x^{i}}_{a_{1}} \rightarrow  \sum_{\substack{a_{2},a_{3},...,a_{k} = 2 \\ a_{i} \neq a_{j}}}^{n} \frac{\mathbf{s}_{a_{1}a_{2}...a_{k}}}{|a_{1}a_{2}...a_{k}|} \frac{\partial}{\partial x^{i}} |a_{1}a_{2}...a_{k}| = 0 , \ \ \ \ \ \  a_{1} \neq 1,
\end{equation}
where we have called $E^{x^{i}}_{a}= \partial_{x^{i}_{a}} \mathcal{S}_{k}$.\footnote{Nevertheless, note that in general for $k>2$ the scattering equations possess singular solutions \cite{cachazo:2019ngv,Cachazo:2019apa} for which some invariants $ |1a_{2}...a_{k}|$ may vanish with $\tau$ and the expansion \eqref{softscatteringeq} is not valid. Here we will assume that such solutions only contribute at subleading orders in the $\tau$ expansion of $m^{(k)}_n$.}

We can now use the CHY formula and the factorization properties just described to isolate the integral over the soft particle in the CHY formula \eqref{CHYm}. Using the canonical ordering $1 \alpha = (12 \cdot \cdot \cdot n)$:
\begin{equation}
\label{mastersoft}
    m^{(k)}_{n}[1 \alpha |1 \alpha] = \mathlarger{\int} d\mu^{(k)}_{n} \ \bigg[ \mathlarger{\mathrm{PT}_{n}^{(k)}[1 \alpha]} \bigg]^{2} \sim \frac{1}{\tau^{k-1}} \mathlarger{\int} d\mu^{(k)}_{n-1} \bigg[ \mathlarger{\mathrm{PT}^{(k)}_{n-1}[\alpha]} \bigg]^{2} S^{(k)}_{n}(1\alpha|1\alpha),
\end{equation}
where
\begin{equation}
\label{ksoftfactor}
      S^{(k)}_{n}(1\alpha|1\alpha) = \int_{} \Bigg[ \frac{d^{k-1}x^{i}_{1}}{ \prod_{i} \hat{E}^{x^{i}}_{1} } \Bigg] \Bigg(  \frac{|n2...k| ... |(n-k+2) (n-k+3)... n2|}{|12\cdot \cdot k||n1 \cdot \cdot k-1| ... |(n-k+2) (n-k+3) \cdot \cdot 1|} \Bigg)^{2},
\end{equation}
and the contour in this soft factor integral is taken to enclose the zeros of the soft scattering equations $|\hat{E}^{x^{i}}_{1}| = \epsilon$ for fixed hard punctures $x^i_a$, $a>1$. This integral can be computed in the standard way, summing over all solutions of the soft scattering equations:
\begin{equation}
\sum_{\hat{E}^{x^{i}}_{1} =0}\Bigg( \frac{|n2...k|...|(n-k+2)(n-k+3)...n|}{|12\cdot\cdot k||n1\cdot\cdot k-1|...|(n-k+2)(n-k+3)\cdot\cdot1|} \Bigg)^{2} \left|\frac{d\hat{E}^{x^{j}}_{1}}{dx^{i}_{1}}\right|^{-1}.
\end{equation}

From now on we will omit the subindex $n$ and the orderings in the soft factor, as it will be implicitly assumed that we are working with identity amplitudes and a $n$-particle scattering process. That is, we write $S^{(k)}_{n}(\mathbb{I}_{n}|\mathbb{I}_{n}) = S^{(k)}$. Given the results we have found for \trg(3,6) in section \ref{section2}, it is natural to expect the following soft factorization:
\begin{equation}
    m^{(k)}_{n}[1 \alpha |1 \alpha] = \frac{1}{\tau^{k-1}} \ \bigg( S^{(k)} m^{(k)}_{n-1}[\alpha | \alpha] + \mathcal{O}(\tau) \bigg).
\end{equation}

In particular, we have just shown, via the CHY representation, that the leading order of the amplitude is $\tau^{-(k-1)}$, e.g. $\tau^{-2}$ for $k=3$. 
The difficulty that one now faces with this relation is that (\ref{ksoftfactor}) apparently depends on the position of the punctures of the hard particles. If this was truly the case, one could not factorize the soft factor out of the $n-1$ particles CHY integral in (\ref{mastersoft}). This translates into the condition for the existence of a soft theorem for tropical Grassmannian amplitudes: if one can show that the integral (\ref{ksoftfactor}) is independent of all the punctures of the hard particles, then a soft theorem is guaranteed to exist, even when the explicit form of the soft factor would not be necessarily known. 

Before closing this subsection we remark an important consequence of our definition of soft limit: It is not respected by the duality \eqref{duality}: From the perspective of \trg$(n-k,n)$ the above limit is equivalent to

\begin{equation}\label{hardresc}
    \mathbf{s}_{b_1\ldots b_{n-k}} = \tau \hat{\mathbf{s}}_{b_1\ldots b_{n-k}}\,,\,\,b_i\neq 1,\quad \tau \to 0,
\end{equation}
which means we can drop $\mathbf{s}_{b_1\ldots b_{n-k}}$ with respect to invariants containing label 1,  $\mathbf{s}_{1b_2\ldots b_{n-k}}$. Thus, this is nothing but a ``hard limit'' of particle 1, already introduced in subsection \ref{subsection2.2} for the \trg{(3,6)} case. Importantly, in this case we have taken the quantities in \eqref{hardresc} as independent up to the condition
\begin{equation}
     \sum_{b_1\ldots b_{n-k} \neq 1}^{n} \mathbf{s}_{b_1\ldots b_{n-k}} = 0,
\end{equation}
which is the condition \eqref{softcon} from the perspective of the dual Grassmannian.

The rest of this paper is dedicated to show that a soft and a hard theorem indeed exists for all tropical Grassmannian amplitudes \trg{($k$,$n$)}, for arbitrary $k$ and $n$. Moreover, we will perform the integral and find the explicit form of the soft factorization. For $k=3$ we show this by direct evaluation of the integral via the GRT. This result paves the way for the general case, which relies also on the ideas presented in section \ref{section4}.

\subsection{Direct Proof of the Soft Theorem for \trg({3,$n$})} \label{subsection{3.3}}

Let us here prove and provide the soft theorem for $k=3$ at all multiplicity via a direct application of the GRT. This will provide the insight that we will build on in section \ref{section4}. In order to make the derivation transparent we will adopt homogeneous Plucker coordinates on $X(3,n)$, which we denote as $(abc)$ instead of $|abc|$. We can then write the soft factor integral \eqref{ksoftfactor} in projective form:
\begin{equation}
\label{projectivesoft}
    S^{(3)} = \mathlarger{\int} \frac{(\sigma d^{2} \sigma)(XY\sigma)}{\Big( \underline{\sum_{b,c} \frac{\mathbf{s}_{1bc} (Xbc)}{(\sigma bc)}} \Big) 
    \Big(\underline{\sum_{b,c} \frac{\mathbf{s}_{1bc} (Ybc)}{(\sigma bc)} } \Big)
    } \bigg(\frac{(n-1,n,2)(n,2,3)}{(n-1,n,\sigma)(n,\sigma,2)(\sigma,2,3)}\bigg)^{2},
\end{equation}
where $X, Y \in \mathbb{C}\mathbb{P}^{2}$ are arbitrary reference vectors, and $\sigma$ is the puncture associated to the soft particle (i.e. particle 1). We will now use underline notation to denote the selected contours. As we have not picked a specific chart of $\mathbb{CP}^2$, the GRT as stated in \cite{griffiths2011principles} will hold without the need of boundary terms, commonly understood as residues at infinity for a given chart.

\begin{figure}
        \centering
        \includegraphics[width=0.6\textwidth]{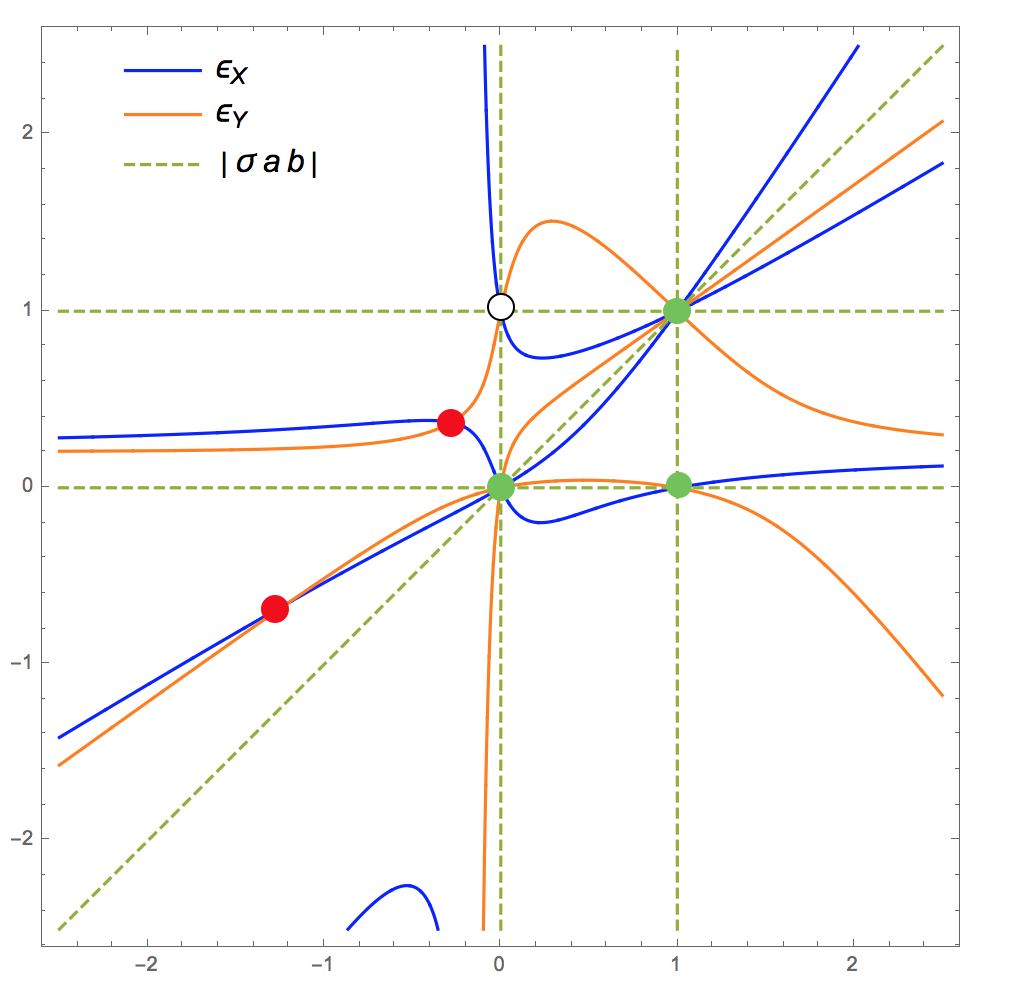} 
        \caption{Real slice with the scattering equations for  $k{=}3,n{=}5$ and generic $X,Y\in \mathbb{CP}^2$, the two solutions are shown as red blobs. Both curves $E_X=0$ and $E_Y=0$ are discontinuous at the white and green blobs: The scattering equations diverge where two lines $|\sigma ab|=0$ intersect but are forced to pass through a neighborhood of these points. On the other hand, the green blobs are included in the locus of $\bar{E}_X$ and $\bar{E}_Y$. The white blob could also be included but will yield no residue, see section \ref{section4}. Hence, via the GRT, the new contour will enclose only the green blobs instead of the red ones.}\label{fig:fig2}
\end{figure}

In order to use the GRT note that the locus of each of the scattering equations is discontinuous where two lines $(\sigma ab)=0$ intersect, see Figure \ref{fig:fig2}. We can soften some of these points by defining $\bar{E}_{X}=(n-1,n,\sigma)(n,\sigma,2)(\sigma,2,3)E_{X}$ and the same for $\bar{E}_{Y}$. The curves $\bar{E}_{X}$ and $\bar{E}_{Y}$ have the same zeroes as the scattering equations, but also include three additional points defined by the pairings $(n-1,n,\sigma)=(n,\sigma,2)=0$, $(n,\sigma,2)=(\sigma,2,3)=0$, and $(n-1,n,\sigma)=(\sigma,2,3)=0$. Note also that $\{\bar{E}_{X},\bar{E}_{Y}\}$ include all possible singularities in the integration and hence can be continued to divisors on $\mathbb{CP}^{2}$, which means we can implement the GRT as stated in \cite{griffiths2011principles}. The statement becomes
\begin{equation}\label{eq:globalrt}
0=\mathlarger{\int}\frac{(\sigma d^{2}\sigma)(XY\sigma)}{\underline{\bar{E}_{X}} \times \underline{\bar{E}_{Y}}}\left((n-1,n,2)(n,2,3)\right)^{2} \,,   
\end{equation}
where we sum over residues associated to solutions of $\bar{E}_X=\bar{E}_Y=0$. This implies $S^{(3)}$ can be computed from the residues at 1) the original zeroes of the scattering equations as in \eqref{projectivesoft} or 2) the three new zeroes defined by the lines $(\sigma ab)$ in the CHY integrand. This generalizes the contour deformation argument reviewed in the introduction for $\mathbb{CP}^1$.

Consider first the contribution coming from the first pairing $(n-1, n, \sigma) = (n, \sigma, 2) = 0$. To evaluate this contribution we change variables
\begin{equation}\label{blowup}
    \sigma = \sigma_n + \epsilon A,
\end{equation}
where $\epsilon$ is one of our new variables. This means the variable $A \in \mathbb{C}\mathbb{P}^{2}$ only carries one independent component. As we will further elaborate in section \ref{section4}, this change of variables makes manifest the fact that puncture 1 is colliding with puncture $n$, as forced by the pairing $(n-1, n, \sigma) = (n, \sigma, 2) = 0$. The corresponding codimension-2 singularity is obtained by $\epsilon \rightarrow 0$, and either $(n-1, n, A) \rightarrow 0$, or $(n, A, 2) \rightarrow 0$. The measure and the integrand change in the following way:

\begin{equation}
    (\sigma d^{2} \sigma) = (n,A,dA) \epsilon d\epsilon, 
\end{equation}
\begin{equation}
    \bigg(\frac{(n-1,n,2)(n,2,3)}{(n-1,n,\sigma)(n,\sigma,2)(\sigma,2,3)}\bigg)^{2} = \bigg(\frac{(n-1,n,2)}{\epsilon^{2} (n-1,n,A)(n,A,2)}\bigg)^{2} + \mathcal{O}\Big( \frac{1}{\epsilon^3} \Big).
\end{equation}

The behaviour of one of the scattering equations as $\epsilon\rightarrow 0$ can be simplified by setting, for instance, $X=n$. This also yields $(XY\sigma)=\epsilon(nYA)$. It is now straightforward to evaluate the $\epsilon \rightarrow 0$ residue,

\begin{equation*}
    \int \frac{(n,A,dA) \epsilon^{3} d\epsilon (n,Y,A)}{\Big[ \sum_{b,c\neq n} \frac{\mathbf{s}_{1bc} (nbc)}{(nbc)} \Big] \Big[\sum_{c} \frac{\mathbf{s}_{1nc} (Ync)}{(Anc)} \Big]} \bigg(\frac{(n-1,n,2)}{\underline{\epsilon^{2}}\, \underline{(n-1,n,A)(n,A,2)}}\bigg)^{2}
\end{equation*}
\begin{equation*}
    = \frac{1}{\mathbf{t}_{2 \ldots n-1}} \int \frac{(n,A,dA)(n,Y,A)}{\Big[ \sum_{c} \frac{\mathbf{s}_{1nc} (Ync)}{(Anc)}\Big]} \bigg(\frac{(n-1,n,2)}{\underline{ (n-1,n,A)(n,A,2)}}\bigg)^{2}
\end{equation*}
\begin{equation}
    = \frac{1}{\mathbf{t}_{2 \ldots n-1}} \int \frac{(A,dA)(Y,A)}{\Big[ \sum_{c} \frac{\mathbf{s}_{1nc} (Yc)}{(Ac)}\Big]} \bigg(\frac{(n-1,2)}{\underline{ (n-1,A)(A,2)}}\bigg)^{2}, \label{eq:lastline}
\end{equation}
where we have defined the planar kinematic invariant $\mathbf{t}_{2 \ldots n-1} = \sum_{a} \mathbf{s}_{1an}$. In the second equality we have also defined the two-bracket $(PQ) \equiv (PnQ)$, since the only role that $n$ plays in such case is to project out one component.

In the last line of \eqref{eq:lastline} we recognize precisely the $k=2$ integral performed in the introduction of this paper. To see this more clearly set $Y=(0,1)$ and pick the chart $A\to (1,\sigma),\sigma_c \to (1,\sigma_c)$ in $\mathbb{CP}^1$. Very nicely, the remaining integral then turns into a $k=2$ soft factor,

\begin{equation*}
    \int \frac{(A,dA)(Y,A)}{\Big[ \sum_{c} \frac{\mathbf{s}_{1nc} (Yc)}{(Ac)}\Big]} \bigg(\frac{(n-1,2)}{ \underline{(n-1,A)(A,2)}}\bigg)^{2}
\end{equation*}
\begin{equation}
     = \int \frac{d \sigma}{\sum_{c} \frac{\mathbf{s}_{1nc}}{\sigma-\sigma_{c}}} \bigg( \frac{\sigma_{n-1} - \sigma_{2}}{\underline{(\sigma_{n-1}-\sigma)(\sigma-\sigma_{2})}} \bigg)^{2} = \frac{1}{\mathbf{s}_{1nn-1}} + \frac{1}{\mathbf{s}_{1n2}}.
\end{equation}

We finish this computation remarking that the last observation regarding a recursive structure in the soft factor will show to be instrumental for our proof of a soft factor at arbitrary $k$.

One can show that the pairing given by $(n,\sigma,2)=(\sigma,2,3)=0$ leads to an answer similar to the last one. The corresponding contribution is
\begin{equation}
    \frac{1}{\mathbf{t}_{3 \ldots n}}\bigg( \frac{1}{\mathbf{s}_{123}} + \frac{1}{\mathbf{s}_{1n2}} \bigg).
\end{equation}

The last contribution is given by the pairing $(n-1,n,\sigma)=(\sigma,2,3)=0$. Geometrically, this contribution is interpreted as the soft puncture standing at intersection point between the lines defined by punctures 2-3, and $(n-1)$-$n$. We call this intersection point $\xi$. Changing variables $\sigma = \alpha \sigma_n + \beta \sigma_2 + \xi$, we note that the two lines are reached as $\alpha \to 0$ and $\beta \to 0$ respectively. We also have
\begin{equation}
    (\sigma d^{2}\sigma) = d\alpha d\beta (2,n,\xi),
\end{equation}
\begin{equation}
    \bigg(\frac{(n-1,n,2)(n,2,3)}{(n-1,n,\sigma)(n,\sigma,2)(\sigma,2,3)}\bigg)^{2} = \bigg(\frac{1}{\alpha \beta (2,n,\xi)}\bigg)^{2}.
\end{equation}
Working out (\ref{projectivesoft}) one can check that the corresponding pole contribution is:
\begin{equation}
    \int\frac{d\alpha d\beta(2n\xi)(XY\sigma)}{\left[\frac{s_{123}(X23)}{\alpha(n23)}+\frac{s_{1n-1n}(Xn-1,n)}{\beta(2,n-1,n)}\right]\left[\frac{s_{123}(Y23)}{\alpha(n23)}+\frac{s_{1n-1n}(Yn-1,n)}{\beta(2,n-1,n)}\right]}\left(\frac{1}{\underline{\alpha}\, \underline{ \beta} (2n\xi)}\right)^{2},
\end{equation}
where again $X, Y \in \mathbb{C}\mathbb{P}^{2}$ are arbitrary reference vectors, which we can take to be $X=\sigma_2$, and $Y=\sigma_n$. It is straightforward to evaluate the integral with this choice, which leads to a contribution 
\begin{equation}
    \frac{1}{\mathbf{s}_{123} \mathbf{s}_{1n-1n}}.
\end{equation}
That is, by direct evaluation of the CHY integral, we have proven the existence of a soft theorem for \trg{(3,$n$)} amplitudes. Adding all the contributions, the $k=3$ soft factor is given by
\begin{equation}
\label{conjecturesoftfactor}
     S^{(3)} = 
    \frac{1}{\hat{\mathbf{s}}_{n-1 n 1}\hat{\mathbf{s}}_{123}} +
    \frac{1}{\hat{\mathbf{t}}_{3 \cdot \cdot n}} \bigg( \frac{1}{\hat{\mathbf{s}}_{n12}} + \frac{1}{\hat{\mathbf{s}}_{123}} \bigg) + 
    \frac{1}{\hat{\mathbf{t}}_{2 \cdot \cdot n-1}} \bigg( \frac{1}{\hat{\mathbf{s}}_{n12}} + \frac{1}{\hat{\mathbf{s}}_{n-1 n 1}} \bigg).
\end{equation}

\section{Geometric Characterization of the Soft Factor} \label{section4}

Before presenting the soft factor for general $k$ let us take a step back and deconstruct the proof of $k=3$ in more geometrical terms. This allows to unwind the singularity structure of the general \trg{($k$,$n$)} soft factor. We study first the case $k=3$, which already shows all the nuances that are present in the general case. We then proceed to the general $k$ case, from which we also discover the general procedure to obtain the soft factor, which we apply in section \ref{section5}. Here we further outline the promised connection between the soft factor and the associahedron $\mathcal{A}_{k-1}$. Note that the connection between ordinary $k=2$ biadjoint theory and the associahedron was introduced in \cite{Mizera:2017cqs}, which also provided a geometric characterization of the corresponding singularities.

\subsection{Singularity Structure} \label{subsection4.1}

\subsubsection{Singularity Structure for $k=3$} \label{subsubsection4.1.1}

The singularities of $m^{(3)}_6$ were characterized in \cite{cachazo:2019ngv} from the geometry of $X(3,6)$. Here we shall follow an analogous strategy to deal only with the soft factorizations, while at the same time considering arbitrary multiplicity. The form of the soft factor for $k=3$ (\ref{conjecturesoftfactor}) tells us directly that the kinematic invariants appear on the soft factor only as simple poles, and in the form of planar combinations. In the following we show that we can recover all such poles as the possible factorization channels associated to the soft scattering equations, and that non-planars contributions must be zero.

Recall the specific integral that we are considering. This time it shall suffice to employ the standard inhomogenous coordinates:
\begin{equation}
\label{geometricsoft}
    S^{(3)} = \int  \frac{dx dy}{(\partial_{x} \mathcal{S}_{3})  (\partial_{y} \mathcal{S}_{3}) }   \bigg[ \frac{|n23||n-1n2|}{|123| |n12| |n-1n1|} \bigg]^{2}.
\end{equation}

We assume that all punctures associated to hard particles are located at generic positions. In section \ref{section3}, we have used the GRT to localize the integration in the singularities of the CHY integrand, rather than in the soft scattering equations. The integrand becomes singular in two different cases. The first case happens when puncture 1 collides with any of the other punctures present in the integrand, namely, punctures 2,3,$n-1$, or $n$. The second case corresponds to the situation in which puncture 1 is collinear with punctures 2 and 3, 2 and $n-1$, or punctures $n-1$ and $n$.\footnote{Other situations (e.g. three particles colliding) do not arise, since our assumption that the hard punctures take generic positions does not allow for such possibility.}

We begin studying the collision of two punctures. Let us take the explicit example of the collision of punctures 1 and 2. To study this collision we perform the change of variables \eqref{blowup}. In inhomogeneous coordinates,
\begin{equation}
\label{collisionzoom}
    x_{1} = x_{2} + \epsilon, \ \ \ \ \ y_{1} = y_{2} + \alpha \epsilon,
\end{equation}
where $\epsilon$ and $\alpha$ correspond to our new variables. We use this coordinate patch to extend the usual notion of blow-up used in $k=2$ to explore geometric singularities, see e.g. \cite{Deligne1969,Brown:2009qja}.

With these variables, the Plucker coordinates become
\begin{equation}
\label{collisiondeterminants}
    |12a| = \epsilon [(y_{2}-y_{a}) + \alpha (x_{a}-x_{2})], \ \ \ \ |1bc| = |2bc| + \mathcal{O}(\epsilon),
\end{equation}
which implies that the potential 1-form $d \mathcal{S}_{3}$ takes the form
\begin{equation}
    d \mathcal{S}_{3} = \frac{d \epsilon}{\epsilon}  \sum_{a=1}^{n} \hat{\mathbf{s}}_{12a} + \mathcal{O}(\epsilon^{0}),
\end{equation}
which corresponds to the factorization channel associated to the planar kinematic invariant $\hat{\mathbf{t}}_{3..n} = \sum_{a} \hat{\mathbf{s}}_{12a}$ in equation (\ref{conjecturesoftfactor}). We now deconstruct our previous CHY computation, now focusing explicitly on how this vertex emerges as a a pole of the soft factor. The change of variables ({\ref{collisionzoom}}) comes along with a non-trivial jacobian:
\begin{equation}
    dx dy = J_{12} d \epsilon d \alpha, \ \ \ \ J_{12} = \epsilon.
\end{equation}
 Another factor of the jacobian $J_{12}$ arises from the localization over the scattering equations, after expressing these in the new coordinates:
\begin{equation}
    \delta^{2}(\partial_{x}S_{3},\partial_{y}S_{3}) \rightarrow J_{12} \delta^{2}(\partial_{\epsilon}S_{3},\partial_{\alpha}S_{3}).
\end{equation}
From (\ref{collisiondeterminants}) we see that the determinants associated to the punctures involved in the collision are linear in $\epsilon$: $|123|, |n12| \sim \epsilon$. An expansion to leading order in $\epsilon$ allows us then to see if the corresponding singularity contributes or not to the integral (\ref{geometricsoft}):
\begin{equation}
     S^{(3)} \sim \mathlarger{\int \frac{J_{12}^{2} d \epsilon}{\frac{\hat{\mathbf{t}}_{3..n}}{\epsilon}} \bigg[ \frac{1}{\epsilon^{2}} \bigg]^{2}} = \frac{1}{\hat{\mathbf{t}}_{3..n}} \int \frac{d \epsilon}{\epsilon},
\end{equation}
from which we see the planar kinematic invariant $\hat{\mathbf{t}}_{3..n}$ contributes as a simple pole. The symbol $\sim$ is taken to mean that we are only paying attention to the $\epsilon$ integral. It is straightforward to show that an analog situation happens when the puncture 1 collides with the puncture $n$.

Following a similar procedure we can deduce that non-planar kinematic invariants cannot appear in the soft factor, since they have a vanishing residue. To exemplify this statement, we can repeat the above construction considering the collision of punctures $1$ and $3$. The main difference with the case presented above for the collision of punctures 1 and 2 is that in the current case only one determinant of order $\epsilon$ makes an appearance in the denominator of (\ref{geometricsoft}). This leads to 
\begin{equation}
    S^{(3)} \sim \mathlarger{\int \frac{\epsilon^{2} d \epsilon}{\frac{\sum \mathbf{s}_{13a}}{\epsilon}} \bigg[ \frac{1}{\epsilon} \bigg]^{2}} = \frac{1}{\sum \mathbf{s}_{13a}} \int \epsilon d \epsilon = 0,
\end{equation}
which shows that there is no contribution associated to the non-planar quantity $\sum_{a} \hat{\mathbf{s}}_{13a}$. It is straightforward to show that an analog situation holds when puncture 1 collides with any puncture different than punctures 2 or $n$ (i.e. those related to 1 by planarity). We conclude that collisions can only give rise to simple poles in the \textit{planar} quantities $\hat{\mathbf{t}}_{3..n}$ and $\hat{\mathbf{t}}_{2..n-1}$.

The only other type of singularity than can arise for $k=3$ is that of a collinear configuration. In order to fix ideas let us consider the limit in which puncture 1 becomes collinear with punctures 2 and 3, and call $\epsilon$ again to the variable that parametrizes how close we are to the singularity. In this case only $|123| \sim \epsilon$, and the potential 1-form is
\begin{equation}
    d\mathcal{S}_{3} = \frac{d \epsilon}{\epsilon}\hat{\mathbf{s}}_{123} + \mathcal{O}(\epsilon^{0}).
\end{equation}

Observe that the collision singularity in moduli space has the  maximal codimension of 2, whereas the collinear configuration is codimension 1. This implies that the jacobian $J_{123}$ corresponding to the blow-up $|123| \sim \epsilon$ scales with a lower power of $\epsilon$ respect to the collision case presented above, i.e. $J_{123} \sim \epsilon^0$. We can now effortlessly repeat the previous procedure, this time for the collinear configuration: We get
\begin{equation}
    S^{(3)} \sim \mathlarger{\int \frac{J_{123}^{2} d \epsilon}{\frac{\mathbf{s}_{123}}{\epsilon}} \bigg[ \frac{1}{\epsilon} \bigg]^{2}} = \frac{1}{\mathbf{s}_{123}} \int \frac{d \epsilon}{\epsilon}.
\end{equation}

An analogous situation happens when puncture 1 is collinear with punctures 2 and $n$, or when it is collinear with punctures $n-1$ and $n$. When puncture 1 is collinear with two other punctures which are not related in the planar ordering, there are no denominators in (\ref{geometricsoft}) providing factors of $\epsilon$ in the denominator. This implies that the contribution for a collinear, non-planar configuration is vanishing. We conclude that collinear limit singularities can only give rise to simple poles in the planar kinematic invariants $\hat{\mathbf{s}}_{123}$, $\hat{\mathbf{s}}_{n12}$, and $\hat{\mathbf{s}}_{n-1n1}$. This exhausts the poles we found in the soft factor \eqref{conjecturesoftfactor}.

\subsubsection{Singularity Structure for Arbitrary $k$} \label{subsubsection4.1.2}

Now we can extend the previous analysis to general $k$ and consider singularities of all codimensions. Let us recall the CHY formula for the arbitrary $k$ soft factor:
\begin{equation}
\label{geometricsoftk}
    S^{(k)} = \int  \bigg[ \frac{d^{k-1}x^{i}}{ \prod_{i} \partial_{x^{i}} \mathcal{S}_{k}} \bigg] \Bigg(  \frac{|n2...k| ... |(n-k+2) (n-k+3)... n2|}{|12\cdot \cdot k||n1 \cdot \cdot k-1| ... |(n-k+2) (n-k+3) \cdot \cdot 1|} \Bigg)^{2}.
\end{equation}

Let us define a degree-$(p-2)$ singularity to the configuration in which puncture 1 tends to the $(p-2)$-plane defined by other $p$-1 punctures. We say that the $p$ punctures become coplanar, in a general higher-dimensional sense. For example, when two punctures collide we are in presence of a situation that we call a degree-0 (or maximal codimension) singularity. The situation in which three punctures become collinear would be a degree-1 singularity.

We now consider the general case of a degree-$(p-2)$ singularity. We are going to show that the ordered coplanar configurations give the precise divergence in \eqref{geometricsoftk} to create a simple pole, it should then be clear that non-planar configurations have less divergence and do not contribute to the soft factor. This is consistent with our application of the GRT: We can deform our contour in an analagous way to the $k=3$ case and we will enclose precisely the planar singularities.

Without any loss of generality we can take the case where 1,2,..., $p$ become coplanar. We change variables in such a way that we zoom-in into the singularity. Again, we call $\epsilon$ the variable that parametrizes the zoom. The $\mathcal{O}(\epsilon)$ determinants together with the corresponding potential 1-form $d\mathcal{S}_{k}$ are:
\begin{equation}
\label{epsilondeterminant}
    |12...p a_{1}...a_{k-p}| \sim \epsilon,
\end{equation}
\begin{equation}
\label{potential1formk}
    d \mathcal{S}_{k} =  \frac{d \epsilon}{\epsilon} \sum_{a_{1},...,a_{k-p}} \hat{\mathbf{s}}_{12...p a_{1}...a_{k-p}} + \mathcal{O}(\epsilon^{0}).
\end{equation}

There are $k-p+1$ brackets in the integrand in (\ref{geometricsoftk}) that are of the form (\ref{epsilondeterminant}), and consequently, the integrand diverges as
\begin{equation}
    \Bigg(  \frac{|n2...k| ... |(n-k+2) (n-k+3)... n2|}{|12\cdot \cdot k||n1 \cdot \cdot k-1| ... |(n-k+2) (n-k+3) \cdot \cdot 1|} \Bigg)^{2}\sim\frac{1}{\epsilon^{2(k-p+1)}}.
\end{equation}

Now, to realize the behaviour \eqref{epsilondeterminant} we need to perform a blow-up as the soft particle approaches the hyperplane defined by the punctures $23\ldots p$. That is, we set $\sigma=X_{2\ldots p}+\epsilon A$, where $\sigma$ is the puncture associated to particle 1, $X_{2\ldots p}$ lies in the plane formed by $23\ldots p$, and
we take $\epsilon \to 0$. The singularity that this generates can be easily read off from (\ref{geometricsoftk}), in appropriate coordinates. Observe that by projectivity $X_{2\ldots p}$ has $p-2$ degrees of freedom, and hence $A$ has $k-p$ degrees of freedom. It is easy to see that the measure becomes
\begin{equation}
    d^{k-1}x^{i}=d^{p-2}Xd^{k-p}A\times\epsilon^{k-p}d\epsilon,
\end{equation}
and a second jacobian factor $\epsilon^{k-p}$ arises due to the localization over the scattering equations:
\begin{equation}
    \int d^{k-1}x^{i}\prod_{i}\delta \Big(\frac{\partial \mathcal{S}_{k}}{\partial x^{i}}\Big)=\int d^{p-2}Xd^{k-p-2}Ad\epsilon\times\delta \Big(\frac{\partial \mathcal{S}_{k}}{\partial\epsilon}\Big)\delta \Big(\frac{\partial \mathcal{S}_{k}}{\partial A}\Big)\delta \Big(\frac{\partial \mathcal{S}_{k}}{\partial X}\Big)\times(\epsilon^{k-p})^{2},
\end{equation}
where $\frac{\partial \mathcal{S}_{k}}{\partial\epsilon}$ is given by (\ref{potential1formk}).






As we approach the residue at $\epsilon \to 0$ in this coordinates the singularity generated by the $\epsilon$-integration is:
\begin{align}
S^{(k)} & \sim\int_{\epsilon\to0}\frac{d\epsilon\,(\epsilon^{k-p})^{2}}{\frac{1}{\epsilon}\sum_{a_{1},...,a_{k-p-2}}\hat{\mathbf{s}}_{12...pa_{1}...a_{k-p-2}}}\frac{1}{\epsilon^{2(k-p+1)}}\nonumber \\
 & =\frac{1}{\sum_{a_{1},...,a_{k-p-2}}\hat{\mathbf{s}}_{12...pa_{1}...a_{k-p-2}}}\label{eq:singeps}.
\end{align}
As this holds for any value of $p$ and any $p$ consecutive labels, we find that the complete set of singularities can be characterized as follows. We know that collision singularities only have two contributions to integration: We can denote them as $(12)$ and $(n,1)$. They give poles of the form $\sum \mathbf{s}_{12a_3\ldots}$ and $\sum \mathbf{s}_{n 1 a_3\ldots }$ respectively. Other collisions $(1a)$ have zero residue and therefore do not contribute to integration. Similarly, collinear singularities of 3 punctures have three contributions with non-zero residue, denoted as $(n-1,n,1),(n,1,2),(1,2,3)$. The associateed invariants are $\sum \mathbf{s}_{n-1,n,1,a_3\ldots}$, $\sum \mathbf{s}_{n,1,2,a_3\ldots}$ and $\sum \mathbf{s}_{123a_3\ldots}$. This pattern repeats until we reach the maximal type of singularity, corresponding to a degree-$(k-2)$ singularity. In this maximal case there are $k$ contributions with non-zero residues and therefore $k$ associated planar invariants.

We conclude that all the poles of the soft factor are given by 
\begin{equation}
\label{singularities}
    \hat{\mathbf{T}}^{(r,q)}_{n} := \sum_{a_{1},..,a_{r} = 1}^{n} \hat{\mathbf{s}}_{12..q a_{1}..a_{r} (n-k+q+r+1)..n-1 n},
\end{equation}
where $0\leq r \leq k-2$, $1 \leq q \leq k-r$. Here $r$ denotes the number of summed indices. For $r=0$ we specialize our notation to:
\begin{equation}
\label{planarS}
    \hat{\mathbf{T}}^{(0,q)}_{n} = \hat{\mathbf{S}}^{(q)}_{n} = \hat{\mathbf{s}}_{12..q (n-k+q+1)...n}, \ \ \ \ \ 1 \leq q \leq k.
\end{equation}

In Appendix \ref{appendixAsub1} we build on the particular result \eqref{eq:singeps} as dictacted by the GRT. We show that, in the strict $\epsilon=0$ limit, the remaining integrals over the variables $X$ and $A$ reduce to smaller CHY integrals that have the same form as (\ref{geometricsoftk}). In a nutshell, the integral over $A$ can be performed as a sum of contributions that arise from the smaller brackets $|\ldots nAp+1\ldots|$, where the punctures associated to particles $2,\ldots p$ drop out because of the zoom. The result of these integrations are of course all the remaining planar invariants containing 1 and associated to colinear configurations of all sizes $p$. As a motivation for the next subsection, let us give here a preview of this recursion. The full contribution of the residue in \eqref{eq:singeps} is indeed given by:
\begin{align}
\label{eq:gf2}
S^{(k)} \longrightarrow & \frac{1}{\hat{\mathbf{T}}_{n}^{(k-p,p)} \hat{\mathbf{T}}_{n}^{(p-3,1)}} S^{(k-p+1)}(p+1\ldots n1)S^{(p-2)}(1\ldots n-k+p-2).
\end{align}

A geometrical picture is starting to emerge from this discussion, clearly associated to the singularities of the CHY integrand in \eqref{eq:singeps}. Indeed, as the reader might have noticed, the picture that we are describing is nothing else than that of the associahedron $\mathcal{A}_{k-1}$ \cite{stasheff1,stasheff2}, and the singularity behaviour $S^{(k)}\to S^{(k-p+1)}\times S^{(p-2)}$
is nothing but the \textit{geometric factorization} as described in e.g. \cite{Arkani-Hamed:2017mur}. A simple way to see this is to consider only the cases $n=k+2$ in (\ref{eq:gf2}), so that the soft factors are identical to biadjoint
amplitudes under Grassmannian duality. Let us now provide some further details on this.

\subsection{Connection with the Associahedron} \label{subsection4.2}

Following the previous ideas, in this subsection we attempt to establish a more detailed connection with the associahedron. What we are after is the relation between moduli and kinematic associahedron unveiled by the ABHY construction \cite{Arkani-Hamed:2017mur}. For this, let us consider a real slice of $X(k,n)$, with coordinates living on  $\mathbb{RP}^{k-1}$. Recall also that we take all but the first puncture to be fixed in the soft limit. In this situation, we conjecture that the region described by 
\begin{equation}
\label{eq:ineq}
|12\cdot\cdot k|>0\,,\quad|n1\cdot\cdot k-1|>0,\ldots,\quad|n-k+2,n-k+3,\ldots1|>0\,,
\end{equation}
is diffeomorphic to the interior of the moduli associahedron $\mathcal{A}_{k-1}$. To see the reason for our claim consider the specific configuration with $n=k+2$, which is generic from the perspective of these inequalities. In this case, via Grassmannian duality, the brackets are dual to $\text{SL}(2,\mathbb{C})$ brackets \cite{cachazo:2019ngv}, which in inhomogeneous coordinates over $\mathbb{RP}^1$ give rise to the inequalities
\begin{equation}
    \sigma_{k+1,k+2}>0,\,\sigma_{k,k+1}>0,\ldots\sigma_{34}>0,\,\sigma_{23}>0.
\end{equation}
In this case, singularities correspond to the collision of punctures in $\mathbb{RP}^{1}$, and the vanishing of simultaneous brackets correspond
to simultaneous collision of a subset of punctures. The blow-ups in
$\epsilon$ precisely correspond to the minimal blow-up scheme introduced
by Deligne-Mumford for the compactifcation of moduli spaces \cite{Deligne1969}.

Further, the singularities of the soft factor that were worked out in subsection \ref{subsection4.1} have a direct interpretation in the associahedron. By looking at the planar kinematic invariants of the form $\hat{\mathbf{T}}^{(0,q)}_{n} = \hat{\mathbf{S}}^{(q)}_{n}$ in (\ref{planarS}), one can see that the corresponding vanishing planar brackets correspond to the codimension-1 boundaries of the associahedron defined by the vanishing of one and only one of the brackets in (\ref{eq:ineq}). The remaining planar invariants $\hat{\mathbf{T}}^{(p,q)}_{n}$ correspond to the case in which more than one bracket in (\ref{eq:ineq}) vanishes simultaneously. Naively, one may think that the associated facet has codimension greater than one. In a similar way as in the ordinary $k=2$, this actually does not happen, and vanishing of more than one bracket still gives codimension-1 boundaries. We can understand this using the previous blow-up change of variables $\sigma_{1} = X + \epsilon A$, associated to the $p-2$ plane defined by $p-1$ punctures forming a planar configuration. That $\sigma_1$ approaches this plane is imposed by the vanishing of $(k-1) - (p-2)$ consecutive brackets in (\ref{eq:ineq}). There are then $p-2$ free parameters given by $X$ that naively describe the singularity. However, as we must only take $\epsilon \rightarrow 0$ to approach the plane, we still need to consider the $k-p$ free parameters given by $A$. In total, we are left with $(p-2) + (k-p) = k-2$ free parameters, and the corresponding singularity is indeed a codimension-1 boundary. In other words, the dimension of the boundary is precisely $k-2$ as expected for the associahedron $\mathcal{A}_{k-1}$. From \eqref{singularities} and its preceding discussion we can also count a total of $\frac{(k-1)(k+2)}{2}$ different kinematic singularities given by all $\mathbf{T}^{(r,p)}_n$, obtained from the same number of singular configurations in moduli space. This is satisfying, since this also corresponds to the number of codimension-1 boundaries in $\mathcal{A}_{k-1}$. 

We have thus identified the boundaries of a certain associahedron both in moduli and kinematic space, together with the factorization properties such as \eqref{eq:gf2}. This suggests that the soft scattering equations (\ref{softscatteringeq}) act as diffeomorphism:\footnote{The proof of the bijectivity/smoothness of the map was given in \cite{Arkani-Hamed:2017mur} for the $k=2$ scattering equations. We do not attempt to repeat these steps here.}
\begin{equation}
    \mathcal{A}_{k-1}(\rm{moduli\,space})\longrightarrow\mathcal{A}^{\rm kin}_{k-1}({\rm kinematic\,space}),
\end{equation}
which in particular means that the associahedra defined in moduli and kinematic space are also related via their canonical forms, as observed first by ABHY \cite{Arkani-Hamed:2017mur}. Now, the canonical form of the kinematic associahedron $\mathcal{A}_{k-1}^{\rm kin}$ (the volume form of the tropical Grassmannian \trg(2,$n$)) is indeed the biadjoint amplitude  \cite{Arkani-Hamed:2017mur,Drummond:2019qjk}! Hence, we can already provide a simple argument for the correspondence of (\ref{geometricsoftk}) with the ordinary amplitudes $m^{(2)}_{k+2}$, somehow alternative to our formal proof. All that we need to show is that the CHY formula (\ref{geometricsoftk}) is the canonical form of $\mathcal{A}^{\rm kin}_{k-1}$ for the vertices we have labelled as $\mathbf{T}^{(r,q)}_n$.

The latter step is achieved by showing that under the map defined by the scattering equations,
\begin{equation}
    \Omega_{k}\longrightarrow S^{(k)},
\end{equation}
where $\Omega_{k}$ is the canonical form of the associahedron (\ref{eq:ineq}). In homogeneous coordinates, and assuming the blown-up boundaries to decouple as in \cite{Arkani-Hamed:2017mur}, it takes the natural form 
\begin{equation}
\label{eq:canform}
\Omega_{k}=d{\rm log}\frac{(1,2\cdot\cdot k)}{(n,1\cdot\cdot k-1)}d{\rm log}\frac{(n,1\cdot\cdot k-1)}{(n-1,n,1\cdot\cdot k-2)}\cdots d{\rm log}\frac{(n-k+3\cdot\cdot1,2)}{(n-k+2,n-k+3\cdot\cdot1)}.
\end{equation}

The derivative operator $d$ acts by definition only on label $1$, and hence the form is projective in all labels. In inhomogenous coordinates, it can be rewritten as
\begin{equation}
    \Omega_{k}=\frac{|n2...k|...|(n-k+1)(n-k+2)...n|}{|12\cdot\cdot k||n1\cdot\cdot k-1|...|(n-k+2)(n-k+3)\cdot\cdot1|}d^{k-1}x^{i},
\end{equation}
which is precisely one power of the soft factor in the integrand of (\ref{geometricsoftk}). We can obtain the whole integrand of the soft factor (\ref{geometricsoftk}) applying the push-forward formalism in the same way as ABHY. The push-forward to kinematic space leads to
\begin{equation*}
    \Omega_{k} \longrightarrow
\end{equation*}
\begin{equation}
\label{eq:pshfwd}
\sum_{{\rm sols.}\hat{E}_{i}=0}\frac{|n2...k|...|(n-k+1)(n-k+2)...n|}{|12\cdot\cdot k||n1\cdot\cdot k-1|...|(n-k+2)(n-k+3)\cdot\cdot1|}\left|\frac{dx^{i}}{d\hat{E}_{j}}\right|\left|\frac{d\hat{E}_{j}}{dX^{I}}\right|d^{k-1}X^{I}.
\end{equation}

The $k-1$ variables $X^{I}$ are defined to be planar and parametrize the
kinematic associahedron. We can take them as the following subset of kinematic invariants:
\begin{equation}
    X^{I}=\{\hat{s}_{12\cdot\cdot k},\hat{s}_{n1\cdot\cdot k-1},...,\hat{s}_{n-k+3\cdot\cdot1,2}\} = \{\hat{\mathbf{S}}^{(k)},\hat{\mathbf{S}}^{(k-1)},...,\\ \hat{\mathbf{S}}^{(1)}\}.
\end{equation}
This choice mimics the singularities of the canonical form
(\ref{eq:canform}), except for the fact that we need to solve for $\hat{s}_{n-k+2,n-k+3\cdot\cdot1}$ in terms of the other kinematic invariants due to momentum conservation. One can check that the jacobian between the scattering equations and the planar variables $X^{I}$ is given by

\begin{equation}
    \left|\frac{d\hat{E}_{j}}{dX^{I}}\right|=\frac{|n2...k|...|(n-k+1)(n-k+2)...n|}{|12\cdot\cdot k||n1\cdot\cdot k-1|...|(n-k+2)(n-k+3)\cdot\cdot1|}.
\end{equation}
Plugging this back in (\ref{eq:pshfwd}) and identifying $\left|\frac{dx^{i}}{d\hat{E}_{j}}\right|$ with the CHY jacobian \cite{cachazo:2019ngv}, we see that we recover the CHY expression for the soft factor (\ref{geometricsoftk}) as a sum over all the solutions of the scattering equations. Nevertheless, this time we have obtained this expression coming from the push-forward of the canonical form of the Associahedron $\Omega_{k}$. That is, the canonical form in kinematic space/biadjoint amplitude.

\section{The Soft Factor for \trg($k$,$n$) from $m^{(2)}_{k+2}$ Amplitudes} \label{section5}

Having outlined the connection between the soft factor and the associahedron, the relation to ordinary $k=2$ biadjoint amplitudes should be by now intuitive. In this section we give the precise form of the soft factor using this insight and also outline the formal proof given mostly in Appendix \ref{appendixA}. Note that because of the duality between the kinematic associahedron $\mathcal{A}^{\rm kin}_{k-1}$ and the tropical Grassmannian \trg$(2,k+2)$ the picture we will describe is precisely that of section \ref{section2}, extended to all $k,n$.

Let us begin with a simple observation. It is straightforward to see that for arbitrary $k$, the $n=k+1$ amplitude is equal to one as there are no scattering equations left where to localize the CHY integral. This hints that the soft factor for $n=k+2$ is the amplitude itself. Under Grassmannian duality, the $m^{(k)}_{k+2}$ amplitude can be expressed as a $m^{(2)}_{k+2}$ amplitude.

We first make explicit the relation between the \trg{(3,$n$)} soft factor already derived and $m^{(2)}_5$ amplitudes. It is known from the standard biadjoint scalar theory that

\begin{equation}
    m^{(2)}_{5}(\mathbb{I}_{5},\mathbb{I}_{5}) = \frac{1}{s_{12}s_{34}} + \frac{1}{s_{23}s_{45}} + \frac{1}{s_{34}s_{51}} + \frac{1}{s_{45}s_{12}} + \frac{1}{s_{51}s_{23}}.
\end{equation}
Under duality, $s_{12} \rightarrow \mathbf{s}_{345}$, $s_{23} \rightarrow \mathbf{s}_{145}$, $s_{34} \rightarrow \mathbf{s}_{125}$, $s_{45} \rightarrow \mathbf{s}_{123}$, $s_{51} \rightarrow \mathbf{s}_{234}$, from which we obtain the dual amplitude:
\begin{equation}\label{m5mod}
    m^{(3)}_{5}(\mathbb{I}_{5},\mathbb{I}_{5}) = \frac{1}{\hat{\mathbf{s}}_{4 5 1}\hat{\mathbf{s}}_{123}} +
    \frac{1}{\hat{\mathbf{s}}_{3 4 5}} \bigg( \frac{1}{\hat{\mathbf{s}}_{512}} + \frac{1}{\hat{\mathbf{s}}_{123}} \bigg) + 
    \frac{1}{\hat{\mathbf{s}}_{2 3 4}} \bigg( \frac{1}{\hat{\mathbf{s}}_{512}} + \frac{1}{\hat{\mathbf{s}}_{4 5 1}} \bigg).
\end{equation}
As expected, this amplitude precisely agrees with the soft factor (\ref{conjecturesoftfactor}). This is because for $k=3$, $m^{(3)}_{4}(\mathbb{I}_{4},\mathbb{I}_{4}) = 1$ and the $\tau$-scaling is indeed homogeneous. Considering the soft factor (\ref{conjecturesoftfactor}), we see that it is just a slight modification from \eqref{m5mod} when we take higher values of $n$, with $k=3$ fixed. In more specific words, the latter is evaluated at a shifted value of the kinematic invariants:
\begin{equation*}
    S^{(3)}_{n}(\mathbb{I}|\mathbb{I}) = m^{(3)}_{5}(\mathbb{I}_{5},\mathbb{I}_{5}) \Big( \hat{\mathbf{s}}_{2 3 4} \rightarrow \hat{\mathbf{t}}_{2 \cdot \cdot n-1}, \hat{\mathbf{s}}_{3 4 5} \rightarrow \hat{\mathbf{t}}_{3 \cdot \cdot n}, \hat{\mathbf{s}}_{4 5 1} \rightarrow \hat{\mathbf{s}}_{n-1 n 1}, \hat{\mathbf{s}}_{512} \rightarrow \hat{\mathbf{s}}_{n12} \Big)
\end{equation*}
\begin{equation*}
    = \frac{1}{\hat{\mathbf{s}}_{n-1 n 1}\hat{\mathbf{s}}_{123}} +
    \frac{1}{\hat{\mathbf{t}}_{3 \cdot \cdot n}} \bigg( \frac{1}{\hat{\mathbf{s}}_{n12}} + \frac{1}{\hat{\mathbf{s}}_{123}} \bigg) + 
    \frac{1}{\hat{\mathbf{t}}_{2 \cdot \cdot n-1}} \bigg( \frac{1}{\hat{\mathbf{s}}_{n12}} + \frac{1}{\hat{\mathbf{s}}_{n-1 n 1}} \bigg)
\end{equation*}
\begin{equation}
\label{shiftedk3}
    \! \! \! \! \! \! \! \! \! \! \! \!  \! \! \! \! \! \! = \frac{1}{\hat{\mathbf{S}}^{(1)}\hat{\mathbf{S}}^{(3)}}+
    \frac{1}{\hat{\mathbf{T}}^{(1,2)}} \bigg( \frac{1}{\hat{\mathbf{S}}^{(2)}} + \frac{1}{\hat{\mathbf{S}}^{(3)}} \bigg) + 
    \frac{1}{\hat{\mathbf{T}}^{(1,1)}} \bigg( \frac{1}{\hat{\mathbf{S}}^{(2)}} + \frac{1}{\hat{\mathbf{S}}^{(1)}} \bigg),
\end{equation}
where in the last line we have made use of the variables (\ref{singularities}) to stress the fact that these are indeed the poles of the soft factor.

The usefulness of this result is twofold. On the one hand, as we stressed in section \ref{section2} it yields an insight into the geometrical interpretation of the soft limit, and on the other hand, it very directly hints the form that a soft factor for arbitrary $k$ could take.

\subsection{Statement of the Soft Theorem} \label{subsection5.1}

Our proof of the $k=3$ case, given in section \ref{section3} using the GRT, can be extended into a recursion formula in $k$ that finally leads to $S^{(k)}$ in closed form. We present and provide the proof for this recursion formula in Appendix \ref{appendixAsub1}. The result is:
\begin{equation}
\label{eq:sum}
\boxed{S^{(k)}(1\ldots n)=\sum_{p=2}^{k+1}S_{p}^{(k)}(1\ldots n)}
\end{equation}

The different terms in the previous sum are given as follows. For $2<p<k+1$:
\begin{equation*}
S_{p}^{(k)}(1\ldots n) =\frac{1}{\hat{\mathbf{T}}_{n}^{(k-p,p)}\hat{\mathbf{T}}_{n}^{(p-3,1)}} \bigg( S^{(k-p+1)}(p+1\ldots n1)(s_{1a_{1}\ldots a_{k-p}}\to s_{12\ldots p,a_{1}\cdots a_{k-p}}) \bigg)
\end{equation*}
\begin{equation}
\label{pmain}
     \times S^{(p-2)}(1\ldots n-k+p-2)(s_{1a_{1}\ldots a_{p-3}}\to s_{n-k+p-1,\ldots,n,1,a_{1}\cdots a_{p-3}}) \bigg).
\end{equation}

Meanwhile, for $p=2$ and $p=k+2$:
\begin{align}
\label{p2}
& S_{2}^{(k)}(1\ldots n) =\frac{1}{\hat{\mathbf{T}}_{n}^{(k-2,2)}}\times S^{(k-1)}(3\ldots n1)(s_{1a_{1}\ldots a_{k-2}}\to s_{12,a_{1}\cdots a_{k-2}}),
\\
\label{pk1}
& S_{k+1}^{(k)}(1\ldots n) =\frac{1}{\hat{\mathbf{T}}_{n}^{(k-2,1)}}\times S^{(k-1)}(1\ldots n-1)(s_{1a_{1}\ldots a_{k-2}}\to s_{n1,a_{1}\cdots a_{k-2}}).
\end{align}

It is straightforward to see that we recover the correct $k=2$ soft factor when we take $S^{(1)} = 1$ as the seed of our recursion relation. For instance, for the case $k=3$:
\begin{equation}
S^{(3)}(1\ldots n)=\sum_{p=2}^{4}S_{p}^{(3)}(1\ldots n).
\end{equation}

When $p = 3$, (\ref{pmain}) gives a contribution
\begin{equation}
    S^{(3)}_{3}(1 \ldots n) = \frac{1}{\mathbf{s}_{123} \mathbf{s}_{1 n-1 n}},
\end{equation}
while the $p=2$ and $p=4$ contributions can be read off from (\ref{p2}) and (\ref{pk1}), giving:

\begin{equation*}
S_{2}^{(3)}(1\ldots n) =\frac{1}{\sum \mathbf{s}_{12a_{1}}} S^{(2)}(3\ldots n1)(\mathbf{s}_{1a_{1}}\to \mathbf{s}_{12a_{1}})
\end{equation*}
\begin{equation}
    = \frac{1}{\mathbf{t}_{3 \ldots n}}\bigg( \frac{1}{\mathbf{s}_{123}} + \frac{1}{\mathbf{s}_{12n}}\bigg),
\end{equation}
and

\begin{equation*}
    S_{4}^{(3)}(1\ldots n) =\frac{1}{\sum \mathbf{s}_{n1a_{1}}} S^{(2)}(1\ldots n-1)(\mathbf{s}_{1a_{1}}\to \mathbf{s}_{n1a_{1}}).
\end{equation*}
\begin{equation}
    S_{4}^{(3)}(1\ldots n) =\frac{1}{\mathbf{t}_{2 \ldots n-1}} \bigg( \frac{1}{\mathbf{s}_{12n}} + \frac{1}{\mathbf{s}_{1n-1n}}\bigg).
\end{equation}
Adding these three contributions obtained through the recursion relation, we precisely get the expression (\ref{shiftedk3}) for the $k=3$ soft factor.

The solution for the recursion relation for arbitrary values of $k$ can be given in the grounds of our previous observations. That is, we expect the solution to be a $m^{(k)}_{k+2}(\mathbb{I},\mathbb{I})  \sim m^{(2)}_{k+2}(\mathbb{I},\mathbb{I})$ amplitude, evaluated at planar combinations of kinematic invariants just as in equation (\ref{shiftedk3}). Concretely, the closed form expression that solves the recursion relation is given by:


\begin{equation}
\label{closedform}
    \boxed{S^{(k)} = m^{(k)}_{k+2} ( \mathbb{I}|\mathbb{I}) \Big(\hat{\mathbf{T}}^{(p,q)}_{k+2} \longrightarrow \hat{\mathbf{T}}^{(p,q)}_{n} \Big)}
\end{equation}
where $\hat{\mathbf{T}}^{(p,q)}_{n}$ are the planar kinematic invariants defined in (\ref{singularities}).

This closed formula for the soft factor works because the recursion (\ref{eq:sum}) is nothing else than the recursion relation presented in \cite{Dolan:2013isa} for the planar cubic theory, which amplitudes correspond as well to the identity amplitudes of the biadjoint scalar theory. The latter amplitudes are dual to the $m^{(k)}_{k+2}$ amplitudes proposed here. We provide the proof of this expression in Appendix \ref{appendixAsub2}.

To close the section, we apply formula (\ref{closedform}) to \trg{(4,$n$)} amplitudes. For this specific case, let us rewrite the kinematic invariants $\hat{\mathbf{T}}^{(p,q)}_{n}$ in the following way:
\begin{equation}
    \hat{\mathbf{t}}_{4..n} = \sum_{a}{\hat{\mathbf{s}}_{123a}},  \ \ \ \ \hat{\mathbf{t}}_{3..n-1} = \sum_{a}{\hat{\mathbf{s}}_{12an}}, \ \ \ \
    \hat{\mathbf{t}}_{2..n-2} = \sum_{a}{\hat{\mathbf{s}}_{1an-1n}},
\end{equation}
\begin{equation}
    \hat{\mathbf{u}}_{3...n} = \sum_{a,b}{\hat{\mathbf{s}}_{12ab}}, \ \ \ \ \hat{\mathbf{u}}_{2...n-1} = \sum_{a,b}{\hat{\mathbf{s}}_{1abn}}.
\end{equation}

The soft factor for \trg{(4,$n$)} amplitudes is:
\begin{equation*}
     S^{(4)} = \frac{1}{\hat{\mathbf{u}}_{3...n}}\bigg\{ \frac{1}{\hat{\mathbf{s}}_{12n-1n} \hat{\mathbf{s}}_{1234}} + \frac{1}{\hat{\mathbf{t}}_{3..n-1}} \bigg[ \frac{1}{\hat{\mathbf{s}}_{12n-1n}}  +  \frac{1}{\hat{\mathbf{s}}_{n123}} \bigg] + \frac{1}{\hat{\mathbf{t}}_{4..n}} \bigg[ \frac{1}{\hat{\mathbf{s}}_{1234}}  +  \frac{1}{\hat{\mathbf{s}}_{n123}} \bigg]\bigg\}
\end{equation*}
\begin{equation*}
    + \frac{1}{\hat{\mathbf{u}}_{2...n-1}} \bigg\{ \frac{1}{\hat{\mathbf{s}}_{123n}\hat{\mathbf{s}}_{1n-2n-1n}} + \frac{1}{\hat{\mathbf{t}}_{3..n-1}} \bigg[ \frac{1}{\hat{\mathbf{s}}_{123n}}  +  \frac{1}{\hat{\mathbf{s}}_{12n-1n}} \bigg]  + \frac{1}{\hat{\mathbf{t}}_{2..n-2}} \bigg[ \frac{1}{\hat{\mathbf{s}}_{n-2n-1n1}}  +  \frac{1}{\hat{\mathbf{s}}_{n-1n12}} \bigg] \bigg\}
\end{equation*}
\begin{equation}
\label{softk4}
   + \frac{1}{\hat{\mathbf{s}}_{1n-2n-1n} \hat{\mathbf{t}}_{4..n} }\bigg[ \frac{1}{\hat{\mathbf{s}}_{1234}} + \frac{1}{\hat{\mathbf{s}}_{123n} } \bigg]  + \frac{1}{\hat{\mathbf{s}}_{1234} \hat{\mathbf{t}}_{2..n-2} }\bigg[ \frac{1}{\hat{\mathbf{s}}_{12n-1n}} + \frac{1}{\hat{\mathbf{s}}_{1n-2n-1n} } \bigg].
\end{equation}

We can also apply the shift formula for the \trg{(5,$n$)} soft factor, the result of which is presented in the Appendix \ref{AppendixB}. Using the cluster polytope results of \cite{Drummond:2019qjk}, we have checked numerically that (\ref{shiftedk3}) provides the soft factors of \trg{(3,7)} and \trg{(3,8)} amplitudes, while (\ref{softk4}) and (\ref{shiftedk5}) provide the soft factors of \trg{(4,7)} and \trg{(5,8)} amplitudes respectively. This entails a highly non-trivial check of the relation between CHY formulas and higher-$k$ tropical Grassmannians.

\section{Discussion and Future Directions} \label{section6}

In this work we have studied the soft limit of a generalization of the biadjoint scalar theory that has been conjectured to be related with the tropical Grassmannian \tg$(k,n)$. The theorem was derived through explicit use of the CHY formula defining the generalization of the biadjoint scalar theory. We note that the theorem holds for all generalized amplitudes in the sense that

\begin{equation}
   m^{(k)}_{n+1}[1\alpha|1\beta ] = \,\, S^{(k)}_{1,\alpha|1,\beta} \times \,m^{(k)}_{n}[\alpha|\beta ] + \mathcal{O}\Big( \frac{1}{\tau^{k-2}} \Big) \,, 
\end{equation}
assuming that the intersection rule we provided in section \ref{section2}, namely
\begin{equation}
    m^{(k)}_{n}[\alpha|\beta] = \sum_{\Upsilon \in J(\alpha) \bigcap  J(\beta)} \Upsilon \,,
\end{equation}
holds for all $k,n$. This rule is a natural corolary of the conjectured relation between CHY and the geometric object arising as certain ordered restriction of the tropical Grassmannian, i.e. $\text{Tr\,G}_{\alpha,\beta}(k,n)$.

With the previous assumption the soft theorem presented in this work implies that the soft factors $S^{(k)}_{1,\alpha|1,\beta}$ for  $\text{Tr\,G}_{\alpha,\beta}(k,n)$ amplitudes can be written in terms of ordinary amplitudes $m^{(2)}_{k+2}[1\alpha|1\beta]$ evaluated at specific combinations of generalized kinematic invariants. For the canonical ordering, the latter correspond to the variables $\mathbf{T}_{n}^{(r,q)}$, in direct bijection to boundaries of the associahedron $\mathcal{A}^{\rm kin}_{k-1}$ or vertices of \trg{(2,$k+2$)}.

This last remark is interesting, for suppose we ask the following question: Is there any theory such that its amplitudes can be interpreted, at any multiplicity, as the soft factor for some other theory? We have shown that the answer to such question is affirmative, provided we broaden our concept of ``theory'' beyond ordinary QFT. For instance, in a situation based in what has already been seen in other remarkable contexts such as \cite{Arkani-Hamed:2013jha, Arkani-Hamed:2013kca, Arkani-Hamed:2018rsk, Damgaard:2019ztj, YelleshpurSrikant:2019meu, Lukowski:2019kqi, Langer:2019iuo} we could define a theory in terms of the (e.g. tropical) geometry, rather than in terms of principles of unitarity and locality. On the other hand, from the perspective of $m^{(k)}_n$ amplitudes it would be interesting to understand general factorization for all $k,n$ in geometrical sense. In this direction an instance of geometrical factorization was studied in \cite{Cachazo:2018wvl} via the "delta algebra" $D_{k,n}$ defined for all $k,n$.

An important consequence can be deduced from the soft theorem introduced in this work when considered along with the duality \trg{($k$,$n$)} $\sim$ \trg{($n-k$,$n$)}. This consequence corresponds to the hard theorem that was presented in subsection \ref{subsection2.2} and also in sec. \ref{subsection{3.2}}. In particular, the limit entails that the objects $S^{(k)}$ provided throughout the paper can be thought not only as soft factors, but also as hard factors emerging in the limit defined by (\ref{hardresc}). Schematically and with appropriate variables, this gives
\begin{equation}
m_{n}^{(k)}\longrightarrow\begin{cases}
S^{(k)}\times m_{n-1}^{(k)}+\mathcal{O}(1/\tau^{{k-2}}) & {\rm \text{(Soft Theorem)}}\\
H^{(k)}\times m_{n-1}^{(k-1)} +\mathcal{O}(1/\tau^{{n-k-2}})  & {\rm \text{(Hard Theorem)}}
\end{cases}
\end{equation}
where $S^{(k)}$ is given by $m^{(2)}_{k+2}$ whereas $H^{(k)}$ is given by $m^{(2)}_{n-k+2}$. Note that while the soft theorem gives a ``$k$-preserving'' factorization, the hard theorem is ``$k$-decreasing'' in resemblance to other contexts \cite{ArkaniHamed:2012nw}. Expectedly, for $k=2$ the hard limit does not lead to any non-trivial factorization of the ordinary amplitudes since $H^{(2)}=m^{(2)}_n$: The hard limit (\ref{hardresc}) becomes $s_{ab}\sim \tau$ for $a,b\neq 1$, but ordinary momentum conservation implies that all kinematic invariants tend to zero. Hence these amplitudes scale homogeneously in $\tau$. Note that a non-trivial instance of hard limit in quantum field theories has recently been studied in \cite{Carrasco:2019qwr}. 

The soft theorem that we have presented in this work is only valid at the leading order in the soft expansion. As such, it presents limitations to explore the full structure of the tropical Grassmannian. This structure includes vertices such as $R_{abcdef}$ and its extensions \cite{Cachazo:2019apa}. Even though they are invisible to the soft factor they appear non-trivially in the hard amplitude $m^{(k)}_{n-1}$. Given that identifying the geometry of the general \tg$(k,n)$ remains a current mathematical problem, it would be interesting to apply "soft bootstrap" techniques (see e.g. \cite{Rodina:2018pcb,Low:2019ynd,Elvang:2018dco})  to tackle this challenge. 

In this direction, a natural step to consider is the study of subleading, and next-to-subleading soft theorems in the generalized biadjoint scalar theory. Indeed, we have found that identity amplitudes for the ordinary biadjoint scalar theory satisfy a new next-to-subleading soft theorem. It can be compactly written as
\begin{equation}
\label{exponentialpart-softfactor}
    S^{(2)}(\mathbb{I|\mathbb{I}}) = \frac{e^{\tau \sum^{n-2}_{a=3} \hat{s}_{1a} \frac{\partial}{\partial s_{a n}}}}{\tau \hat{s}_{1n}} + \frac{e^{\tau \sum^{n-2}_{a=3} \hat{s}_{1a} \frac{\partial}{\partial s_{2 a}}}}{\tau \hat{s}_{12}} + \frac{\tau}{2} \sum_{l=3}^{n-2} \hat{s}_{1l}  \Bigg[ \frac{\partial}{\partial s_{2 l}} - \frac{\partial}{\partial s_{l n}} \Bigg]^{2}+ \mathcal{O}(\tau^2).
\end{equation}
This expression contains the subleading soft theorem presented in \cite{Cheung:2017ems} and can be deduced by performing the soft expansion in the CHY formula, as in \cite{Afkhami-Jeddi:2014fia} (choosing as a basis of kinematic invariants the set $\{ s_{ab}, s_{a2}, s_{an}\}$, where $a \in \{ 3,...,n-2 \}$). It would be interesting to see if any of these relations has an analog for the generalized biadjoint scalar theory and a corresponding dual theorem. If affirmative, it would also be interesting to see if any of these subleading theorems has a realization in terms of the tropical Grassmannian geometry. Notice, however, that application of the same procedure is complicated for higher values of $k$, as singular solutions to the scattering equations arise \cite{cachazo:2019ngv,Cachazo:2019apa}. Further scrutiny is necessary in order to find such subleading soft theorems.

As we showed, the soft factors satisfy a recursion relation that matches the one presented in \cite{Dolan:2013isa} for the planar cubic theory (i.e. identity amplitudes for the biadjoint scalar theory).  It would be fascinating to study if the \trg{($k$,$n$)} amplitudes themselves satisfy some kind of recursion formula that reduces to the recursion formula in \cite{Dolan:2013isa} for the ordinary \trg{(2,$n$)} case. Another possibility left for future study is to establish a Berends-Giele like recursion for the generalized biadjoint scalar theory, as the one worked out in \cite{Mafra:2016ltu} for the ordinary biadjoint scalar theory.  This could be possible if a ``self-similarity'' structure is present in the tropical Grassmannians. Such a result would be interesting, as it could allows us to understand better the factorization structure of the generalized amplitudes defined by the $\mathbb{CP}^{k-1}$ extension of the CHY formalism.

\acknowledgments

We thank F. Cachazo, N. Early and S. Mizera for various useful comments on the draft. In particular, we would like to thank S. Mizera for insightful discussions and F. Cachazo for his continuous encouragement during this project. We also thank J. Rojas for help with a numerical check of the $k=3$, $n=7$ soft factor. D.G.S would like to thank the Perimeter-Scholars-International (PSI) program where this project saw light, and the Abdus Salam International Centre for Theoretical Physics, ICTP-SAIFR/IFT-UNESP and FAPESP grant 2016/01343-7 for partial financial support. A.G. thanks MIAPP for kind hospitality while this work was completed. A.G. also acknowledges financial support via Conicyt grant 21151647. This research was supported by the Munich Institute for Astro- and Particle Physics (MIAPP) of the DFG cluster of excellence "Origin and Structure of the Universe". Research at Perimeter Institute is supported by the Government of Canada through the Department of Innovation, Science and Economic Development Canada and by the Province of Ontario through the
Ministry of Research, Innovation and Science.

\appendix

\section{The Soft Factor $S^{(k)}$ from a Recursion Formula in $k$} \label{appendixA}

\subsection{Derivation of the Recursion} \label{appendixAsub1}

In this appendix we use the GRT to prove the recursion relation \eqref{eq:sum}, which we quote here for convenience:
\begin{equation}
\label{eq:aprec}
S^{(k)}(1\ldots n)=\sum_{p=2}^{k+1}S_{p}^{(k)}(1\ldots n)\,,
\end{equation}
where, for $2<p<k+1$:
\begin{equation*}
S_{p}^{(k)}(1\ldots n) =\frac{1}{\sum s_{12\ldots p,a_{1}\cdots a_{k-p}}}\times \bigg( S^{(k-p+1)}(p+1\ldots n1)(s_{1a_{1}\ldots a_{k-p}}\to s_{12\ldots p,a_{1}\cdots a_{k-p}}) \bigg)
\end{equation*}
\begin{equation*}
  \times\frac{1}{\sum s_{n-k+p-1,\ldots n,1,a_{1}\cdots a_{p-3}}}\times \bigg(
\end{equation*}
\begin{equation}
    S^{(p-2)}(1\ldots n-k+p-2)(s_{1a_{1}\ldots a_{p-3}}\to s_{n-k+p-1,\ldots,n,1,a_{1}\cdots a_{p-3}}) \bigg),
\end{equation}
and for $p=2$ and $p=k+2$:
\begin{equation}
S_{2}^{(k)}(1\ldots n) =\frac{1}{\sum s_{12,a_{1}\cdots a_{k-2}}}\times S^{(k-1)}(3\ldots n1)(s_{1a_{1}\ldots a_{k-2}}\to s_{12,a_{1}\cdots a_{k-2}}),
\end{equation}
\begin{equation}
S_{k+1}^{(k)}(1\ldots n) =\frac{1}{\sum s_{n1,a_{1}\cdots a_{k-2}}}\times S^{(k-1)}(1\ldots n-1)(s_{1a_{1}\ldots a_{k-2}}\to s_{n1,a_{1}\cdots a_{k-2}}).
\end{equation}
%
%
%
%
We take $S^{(1)} = 1$, since with this value we recover the correct soft factor for $k=2$ (which can be found by direct computation). We find the explicit solution to the recursion in the next subsection, where we also find that it can be written in terms of the biadjoint amplitudes $m^{(2)}_{k+2}[\mathbb{I}|\mathbb{I}]$.

To prove the recursion formula, we recall the integral representation of the soft factor,
\begin{equation}\label{orsoft}
        S^{(k)} = \int  \bigg[ \frac{d^{k-1}x^{i}}{ \prod_{i} \underline{\hat{E}_{i}}} \bigg] \Bigg(  \frac{|n2...k| ... |(n-k+2) (n-k+3)... n2|}{|12\cdot \cdot k||n1 \cdot \cdot k-1| ... |(n-k+2) (n-k+3) \cdot \cdot 1|} \Bigg)^{2},
\end{equation}
where we have underlined the scattering equations defining the integration contour. The implementation of the GRT is done in analogous manner to section \ref{subsection{3.3}}. In fact, during section \ref{subsection4.2} we have found that the only singularities of the integration, besides the scattering equations, can arise from the consecutive vanishing of Plucker coordinates in the CHY integrand. Let us then assume that the contour can be deformed in the same way as in section \ref{subsection{3.3}} to enclose all these singularities. More precisely, the GRT applied to \eqref{orsoft} will yield \eqref{eq:aprec}, where $p$-th singularity of the integral is given by
\begin{align}
\label{eq:ff}
S_{p}^{(k)} & =\int\frac{d^{k-1}x^{i}}{\prod_{i}\hat{E}_{i}} \Bigg( \frac{|n-k+2\ldots n2|\cdots|n-k-1+p\ldots n2\ldots p-1|}{\underline{|n-k+2\ldots n1|\cdots|n-k-1+p\ldots n12\ldots p-2|}}\times \nonumber \\
 &\frac{|n-k+p\ldots n2\ldots p|}{|n-k+p\ldots n12\ldots p-1|} \times \frac{|n-k+p+1\ldots n2\ldots p+1|\cdots|n2\ldots k|}{\underline{|n-k+p+1\ldots n12\ldots p|\cdots|1\ldots k|}} \Bigg)^{2}.
\end{align}
To compute this residue we observe that the second line of (\ref{eq:ff}) forces the soft particle to lie in the plane generated by particles $23\ldots p$, hence we perform the blow up on this configuration (we could also blow up the complement instead, namely the plane defined
by the first line of (\ref{eq:ff})). We write
\begin{align}
\sigma=X_{2\ldots p}+\epsilon A, 
\end{align}
where $X_{2\ldots p}$ ($X$ for short) has $p-2$ coordinates $\chi^{R}$ and $A$ has $k-p$ coordinates $\alpha^{r}$. Here we assumed $p\leq k$ which excludes the case $p=k+1$ in (\ref{eq:sum}), in which we are forced to blow up the complement plane instead. Now, the analysis
of the $\epsilon$- integration was already outlined in subsection \ref{subsection4.1},
hence we jump straight to the result in the strict $\epsilon=0$ limit. In homogeneous coordinates

\begin{align}
\label{eq:ff2}
S_{p} & =\frac{1}{\sum\hat{s}_{12\ldots p,a_{1}\cdots a_{k-p}}}\int\frac{(AXd^{k-p}Ad^{p-2}X)}{\text{\ensuremath{\prod_{r}}}\frac{\partial S}{\partial\alpha^{r}}\text{\ensuremath{\prod_{R}}}\frac{\partial S}{\partial\chi^{R}}}(AXA^{(1)}\cdots A^{(k-p)}X^{(1)}\cdots X^{(p-2)})\nonumber \\
 & \left(\frac{(n-k+2\ldots n2)\cdots(n-k-1+p\ldots n2\ldots p-1)}{\underline{(n-k+2\ldots nX)\cdots(n-k-1+p\ldots nX2\ldots p-2)}}\times\frac{(n-k+p\ldots n2\ldots p)}{(n-k+p\ldots nX2\ldots p-1)}\right.\nonumber \\
 & \left.\frac{(n-k+p+1\ldots n2\ldots p+1)\cdots(n2\ldots k)}{\underline{(n-k+p+1\ldots nA2\ldots p)\cdots(A\ldots k)}}\right)^{2},
\end{align}
where the vectors are defined as $A^{(r)}=\frac{\partial A}{\partial\alpha^{r}}$
and $X^{(R)}=\frac{\partial X}{\partial\chi^{R}}$, with the latter
living in the plane $23\ldots p$. Now, at strict $\epsilon=0$:

\begin{align}
\frac{\partial S}{\partial\alpha^{r}} & =\sum_{a}\frac{s_{12\ldots p,a_{1}\cdots a_{k-p}}}{(A2\ldots pa_{1}\cdots a_{k-p})}(A^{(r)}2\ldots pa_{1}\cdots a_{k-p})\\
\frac{\partial S}{\partial\chi^{R}} & =\sum_{a}\frac{s_{1,n-k-1+p\ldots n,a_{1}\cdots a_{p-3}}}{(X,n-k-1+p\ldots n,a_{1}\cdots a_{p-3})}(X^{(R)},n-k-1+p\ldots n,a_{1}\cdots a_{p-3})+\ldots
\end{align}
where in $\frac{\partial S}{\partial\chi^{R}}$ we have only written
the terms that will contribute in the contour marked for $X$, given
by the second line of (\ref{eq:ff2}), which forces $X$ to approach
the plane $n-k-1+p\ldots n$. Note that this fixes $X$ to be the
point in the intersection of such $(k-p+2)$-plane and the $(p-1)$-plane
$2\ldots p$. Also note that the sum over $a$ in the second line
is not constrained by the fact that $X^{(R)}$ is in $2\ldots p$
as the labels $a_{1}\ldots a_{p-3}$ will never contain such plane.

As the plane $2\ldots p$ is spanned by $X,X^{(1)},\ldots,X^{(p-2)}$
we can use the following Schouten identity:
\begin{align}
(AXd^{k-p}Ad^{p-2}X)(AXA^{(1)}\cdots A^{(k-p)}X^{(1)}\cdots X^{(p-2)})(n-k+p\ldots n2\ldots p)^{2} \nonumber \\
=(Ad^{k-p}A,2\ldots p)(AA^{(1)}\cdots A^{(k-p)},2\ldots p) \nonumber \\
\times(n-k+p\ldots nXX^{(1)}\cdots X^{(p-2)})(n-k+p\ldots nXd^{p-2}X).
\end{align}

(to see this identity we can write $dX=d\chi_{R}X^{(R)}$). We can
now factor the integral in $S_{p}^{(k)}$ in two parts. In order to
write them let us define the short brackets
\begin{align}
[T_{1}..\ldots T_{p-1}] & :=(n-k+p\ldots n,T_{1}..\ldots T_{p-1}), \\
\{T_{1}\ldots T_{k-p+1}\} & :=(T_{1}\ldots T_{k-p+1},2\ldots p).
\end{align}
The first integral is over $X$ and can be now written

\begin{align}
& \int\frac{[Xd^{p-2}X]}{\prod_{R}\sum_{a}\frac{s_{1,n-k-1+p\ldots n,a_{1}\ldots a_{p-3}}}{[X,n-k-1+p,a_{1}\cdots a_{p-3}]}[X^{(R)},n-k-1+p,a_{i_{1}}\cdots a_{i_{p-3}}]}[XX^{(1)}\cdots X^{(p-2)}]\times \nonumber\\
& \left[\frac{[n-k+2\ldots n-k+p-1,2]\cdots[n-k-1+p2\ldots p-1]}{\underline{[n-k+2\ldots n-k+p-1X]\cdots[n-k-1+pX2\ldots p-2]}[X2\ldots p-1]}\right]^{2},
\end{align}
which then becomes 
\begin{align}
    S_{p}^{(p-1)}(12\ldots n-k+p-1)(s_{1a_{1}\ldots a_{p-2}}\to s_{1n-k+p,\ldots,n,a_{1}\ldots a_{p-2}}),
\end{align}
with the first argument means this piece must be evaluated for the canonical ordering with particles $n-k+p,\ldots,n$ removed and the second argument indicates a replacement. Observe that as the residue is localized over ``planar'' brackets (w.r.t the aforementioned ordering) only planar invariants among $s_{1n-k+p,\ldots,n,a_{1}\ldots a_{p-2}}$ will appear in $S_{p}^{(p-1)}$. Hence there is no need to impose momentum conservation on them since they are regarded as independent.

The second integral is over $A$ and reads
\begin{align}
\int &\frac{\{Ad^{k-p}A\}\{AA^{(1)}\cdots A^{(k-p)}\}}{\prod_{r}\sum_{a}\frac{s_{12\ldots p,a_{1}\cdots a_{k-p}}}{\{Aa_{1}\cdots a_{k-p}\}}\{A^{(r)}a_{1}\cdots a_{k-p}\}} \times \nonumber \\ & \left(\frac{\{n-k+p+1\ldots n,p+1\}\cdots\{n,p+1\ldots k\}}{\underline{\{n-k+p+1\ldots nA\}\cdots\{Ap+1\ldots k\}}}\right)^{2},
\end{align}
which gives the full soft factor
\begin{align}
    S^{(k-p+1)}(p+1\ldots n1)(s_{1a_{1}\ldots a_{k-p}}\to s_{12\ldots p,a_{1}\cdots a_{k-p}}),
\end{align}
where particles $2,\ldots,p$ are removed from the ordering. 

We can write our full result as
\begin{align}
S_{p}^{(k)}(1\ldots n) & =\frac{1}{\sum s_{12\ldots p,a_{1}\cdots a_{k-p}}}\times S^{(k-p+1)}(p+1\ldots n1)(s_{1a_{1}\ldots a_{k-p}}\to s_{12\ldots p,a_{1}\cdots a_{k-p}})\nonumber \\
 & \times S_{p}^{(p-1)}(12\ldots n-k+p-1)(s_{1a_{1}\ldots a_{p-2}}\to s_{1n-k+p,\ldots,n,a_{1}\ldots a_{p-2}})\label{eq:pf3}.
\end{align}
 
Recall that this holds for $k\geq p$. Let us provide now an expression for $S_{p}^{(p-1)}$. Specializing (\ref{eq:pf3}) to $p=2$ (recalling
that $S_{p}^{(1)}=1$) we get
\begin{align}
\label{p=2}
S_{2}^{(k)}(1\ldots n)=\frac{1}{\sum s_{12,a_{1}\cdots a_{k-2}}}\times S^{(k-1)}(3\ldots n1)(s_{1a_{1}\ldots a_{k-2}}\to s_{12,a_{1}\cdots a_{k-2}}),
\end{align}
but we can revert the canonical ordering in the definition (\ref{eq:ff})
to write this as
\begin{align}
S_{k+1}^{(k)}(1\ldots n)=\frac{1}{\sum s_{n1,a_{1}\cdots a_{k-2}}}\times S^{(k-1)}(1\ldots n-1)(s_{1a_{1}\ldots a_{k-2}}\to s_{n1,a_{1}\cdots a_{k-2}}),
\end{align}
which implies 
\begin{align}
S_{p}^{(p-1)}&(12\ldots n-k+p-1)=\frac{1}{\sum s_{n-k+p-1,1,a_{1}\cdots a_{p-3}}}\times  \nonumber \\ 
& S^{(p-2)}(1\ldots n-k+p-2)(s_{1a_{1}\ldots a_{p-3}}\to s_{n-k+p-1,1,a_{1}\cdots a_{p-3}}).
\end{align}.

Using this equation, together with equations (\ref{p=2}), (\ref{eq:pf3}), and shifting as indicated, we finally obtain the recursion relation (\ref{eq:aprec}).

\subsection{Solving the Recursion: The Soft Factor as a Biadjoint Amplitude} \label{appendixAsub2}

In this subsection we use the recursion relation in \cite{Dolan:2013isa} for planar cubic theory amplitudes, which coincide with the identity amplitudes for the biadjoint scalar theory. We need the recursion for the specific case $n=k+2$:
\begin{align}
\label{DGrecursion}
    A_{k+2}(k_{1},...,k_{k+2}) & = \frac{1}{(k_{3} + \ldots + k_{k+2})^{2}}A_{k+1}(k_{1}+k_{2},k_{3},\ldots,k_{k+2}) \nonumber \\
    &+ \frac{1}{(k_{2} + \ldots +  k_{k+1})^{2}}A_{k+1}(k_{k+2}+k_{1},k_{2},\ldots,k_{k+1}) \nonumber \\
    &+ \sum_{p=3}^{k} \frac{A_{p}(\pi_{p},k_{2},\ldots,k_{p}) A_{k-p+3}(\bar{\pi}_{p},k_{p+1},\ldots,k_{k+2})}{(k_{2} + \ldots + k_{p})^{2}(k_{p+1} + \ldots + k_{k+2})^{2}},
\end{align}
where
\begin{equation}
    \pi_{m} = -k_{2}-\ldots-k_{m}, \ \ \ \ \ \  \bar{\pi}_{m} = -k_{m}-\ldots-k_{k+2},
\end{equation}
where $k_{i}$ correspond to the momentum of the $i$-th particle.

It is straightforward to see that under duality the propagators transform as:
\begin{equation}
    (k_{3} + \ldots + k_{k+2})^{2} = \sum_{\substack{\mathrm{pairs} (a,b) \\ \mathrm \in \{3,..,k+2\}}} s_{ab} \longrightarrow \sum_{a_{1}, \ldots, a_{k-2} = 1}^{k+2} \mathbf{s}_{12a_{1}...a_{k-2}} = \hat{\mathbf{T}}^{(k-2,2)}_{k+2},
\end{equation}
where the variables $\hat{\mathbf{T}}^{(a,b)}_{n}$ are defined as in section \ref{section5}:
\begin{equation}
    \hat{\mathbf{T}}^{(p,q)}_{n} = \sum_{a_{1},..,a_{p} = 1}^{n} \hat{\mathbf{s}}_{12..q a_{1}..a_{p} (n-k+q+p+1)..n-1 n}. 
\end{equation}
Similarly one can check that:
\begin{align}
    & (k_{2} + \ldots +  k_{k+1})^{2} \longrightarrow \hat{\mathbf{T}}^{(k-2,1)}_{k+2}, \\
    & (k_{p+1} + \ldots + k_{k+2})^{2} \longrightarrow \hat{\mathbf{T}}^{(k-p,p)}_{k+2}, \\
    & (k_{2} + \ldots + k_{p})^{2} \longrightarrow \hat{\mathbf{T}}^{(p-3,1)}_{k+2}.
\end{align}

Let us study now how the amplitudes factors behave under duality. Take, for example, the $A_{k+1}(k_{1}+k_{2},k_{3},\ldots,k_{k+2})$ term. By momentum conservation $k_{1} + k_{2} = - k_{3} - \ldots -k_{k+2}$, and therefore the amplitude can be regarded as a function $A_{k+1}(s_{a_{1}...a_{m}})$ of planar kinematic invariants only, with the indices satisfying a planar order and in the set $\{3,\ldots,k+2\}$. Under duality we get the same amplitude, but evaluated in the corresponding dual kinematic invariants:
\begin{equation}
    A_{k+1}(k_{1}+k_{2},k_{3},\ldots,k_{k+2}) = A_{k+1}\big( s_{a_{1}...a_{m}} \big) \longrightarrow  A_{k+1}\Big(\hat{\mathbf{T}}^{(m-2,a_{1}-1)}_{k+2}\Big).
\end{equation}
We can now follow the same procedure over all amplitude factors appearing in (\ref{DGrecursion}). We first express them in terms of the planar basis of kinematic invariants, and therefore under duality they can all be written in terms of the variables $\hat{\mathbf{T}}^{(a,b)}_{k+2}$:
\begin{equation}
    A_{k+1}(k_{k+2}+k_{1},k_{2},\ldots,k_{k+1})= A_{k+1}\big( s_{a_{1}...a_{m}} \big)  \longrightarrow  A_{k+1}\Big( \hat{\mathbf{T}}^{(m-2,a_{1}-1)}_{k+2} \Big),
\end{equation}
\begin{equation}
    A_{k-p+3}(\bar{\pi}_{p},k_{p+1},\ldots,k_{k+2}) = A_{k-p+3}\big( s_{a_{1}...a_{m}} \big)  \longrightarrow  A_{k+1} \Big( \hat{\mathbf{T}}^{(m-2,a_{1}-1)}_{k+2} \Big),
\end{equation}
\begin{equation}
    A_{p}(\pi_{p},k_{2},\ldots,k_{p}) = A_{p}\big( s_{a_{1}...a_{m}} \big) \longrightarrow  A_{k+1}\Big( \hat{\mathbf{T}}^{(m-2,a_{1}-1)}_{k+2} \Big),
\end{equation}
where in the previous we have used a condensed notation $A_{n}\big( s_{a_{1}...a_{m}}\big)$ to mean $A_{n}$ as a function of the planar kinematic invariants $s_{a_{1}...a_{m}}$ that can be constructed from the momenta appearing in the corresponding left hand side.

The recursion relation (\ref{eq:aprec}) for the soft factor takes the same form as the dualized version of (\ref{DGrecursion}), up to one modification: the summation indices get promoted from the set $\{1,...,k+2\}$ to the set $\{1,...,n\}$. That is, we must promote
\begin{equation}
\label{promotion}
    \hat{\mathbf{T}}^{(a,b)}_{k+2} \longrightarrow  \hat{\mathbf{T}}^{(a,b)}_{n},
\end{equation}
in all instances where a $\hat{\mathbf{T}}^{(a,b)}_{k+2}$ appears. We conclude that the soft factor satisfies the same recursion relation as (\ref{DGrecursion}), considering duality and the replacements (\ref{promotion}). 

We know by direct computation that the base case 
\begin{equation}
S^{(2)} = \bigg( \frac{1}{\hat{s}_{n1}} + \frac{1}{\hat{s}_{12}} \bigg),
\end{equation}
satisfies (\ref{closedform}). Since both expressions in (\ref{closedform}) are equal in the base case $k=2$, and satisfy the same recursion relation, by induction we can conclude that (\ref{closedform}) must be true:
\begin{equation}
    \boxed{S^{(k)} = m^{(k)}_{k+2} ( \mathbb{I}|\mathbb{I}) \Big(\hat{\mathbf{T}}^{(p,q)}_{k+2} \longrightarrow \hat{\mathbf{T}}^{(p,q)}_{n} \Big)}
\end{equation}

\section{Explicit Soft Factor for $k=5$} \label{AppendixB}

In this appendix we provide the explicit form for the \trg{(5,$n$)} soft factor as obtained from the $m^{(k)}_{k+2}$ amplitude. For $n=8$, we have checked this expression against the amplitude provided in \cite{Drummond:2019qjk}.

Let us define
\begin{equation}
    \hat{\mathbf{t}}_{5..n} = \sum{\hat{\mathbf{s}}_{1234a}} \ \ \ \
    \hat{\mathbf{t}}_{4..n-1} = \sum{\hat{\mathbf{s}}_{123an}}
\end{equation}
\begin{equation}
    \hat{\mathbf{t}}_{3..n-2} = \sum{\hat{\mathbf{s}}_{12an-1n}} \ \ \ \    \hat{\mathbf{t}}_{2..n-3} = \sum{\hat{\mathbf{s}}_{1an-2n-1n}}
\end{equation}
\begin{equation}
    \hat{\mathbf{u}}_{4...n} = \sum{\hat{\mathbf{s}}_{123ab}} \ \ \ \ \hat{\mathbf{u}}_{3...n-1} = \sum{\hat{\mathbf{s}}_{12abn}} \ \ \ \ \hat{\mathbf{u}}_{2...n-2} = \sum{\hat{\mathbf{s}}_{1abn-1n}}
\end{equation}
\begin{equation}
    \hat{\mathbf{v}}_{3...n} = \sum{\hat{\mathbf{s}}_{12abc}} \ \ \ \ \hat{\mathbf{v}}_{2...n-1} = \sum{\hat{\mathbf{s}}_{1abcn}}
\end{equation}

The \trg{(5,$n$)} soft factor is given by:

\begin{equation*}
    S^{(5)}_{n}(\mathbb{I}|\mathbb{I}) = \frac{1}{\hat{\mathbf{v}}_{3...n}} \Bigg\{ \frac{1}{\hat{\mathbf{u}}_{4...n}}\bigg[ \frac{1}{\hat{\mathbf{s}}_{12345} \hat{\mathbf{s}}_{123n-1n}}+\frac{1}{\hat{\mathbf{t}}_{5..n}} \bigg( \frac{1}{\hat{\mathbf{s}}_{12345}} + \frac{1}{\hat{\mathbf{s}}_{1234n}} \bigg) 
\end{equation*}
\begin{equation*}
    +\frac{1}{\hat{\mathbf{t}}_{4..n-1}}\bigg( \frac{1}{\hat{\mathbf{s}}_{1234n}} + \frac{1}{\hat{\mathbf{s}}_{123n-1n}} \bigg) \bigg] + \frac{1}{\hat{\mathbf{u}}_{3...n-1}}\bigg[ \frac{1}{\hat{\mathbf{s}}_{1234n} \hat{\mathbf{s}}_{12n-2n-1n}}
\end{equation*}
\begin{equation*}
    +\frac{1}{\hat{\mathbf{t}}_{4..n-1}} \bigg( \frac{1}{\hat{\mathbf{s}}_{1234n}} + \frac{1}{\hat{\mathbf{s}}_{123n-1n}} \bigg) + \frac{1}{\hat{\mathbf{t}}_{3..n-2}} \bigg( \frac{1}{\hat{\mathbf{s}}_{123n-1n}} + \frac{1}{\hat{\mathbf{s}}_{12n-2n-1n}}\bigg) \bigg]
\end{equation*}
\begin{equation*}
    + \frac{1}{\hat{\mathbf{s}}_{12n-2n-1n}\hat{\mathbf{t}}_{5..n}} \bigg( \frac{1}{\hat{\mathbf{s}}_{12345}} + \frac{1}{\hat{\mathbf{s}}_{1234n}} \bigg) + \frac{1}{\hat{\mathbf{s}}_{12345}\hat{\mathbf{t}}_{3..n-2}} \bigg( \frac{1}{\hat{\mathbf{s}}_{123n-1n}} + \frac{1}{\hat{\mathbf{s}}_{12n-2n-1n}} \bigg) \Bigg\}
\end{equation*}
\begin{equation*}
    + \frac{1}{\hat{\mathbf{v}}_{2...n-1}} \Bigg\{ \frac{1}{\hat{\mathbf{u}}_{2...n-2}}\bigg[ \frac{1}{\hat{\mathbf{s}}_{123n-1n} \hat{\mathbf{s}}_{1n-3n-2n-1n}}+\frac{1}{\hat{\mathbf{t}}_{3..n-2}} \bigg( \frac{1}{\hat{\mathbf{s}}_{1n-3n-2n-1n}} + \frac{1}{\hat{\mathbf{s}}_{123n-1n}} \bigg) 
\end{equation*}
\begin{equation*}
    +\frac{1}{\hat{\mathbf{t}}_{3..n-2}}\bigg( \frac{1}{\hat{\mathbf{s}}_{12n-2n-1n}} + \frac{1}{\hat{\mathbf{s}}_{123n-1n}} \bigg) \bigg] + \frac{1}{\hat{\mathbf{u}}_{3...n-1}}\bigg[ \frac{1}{\hat{\mathbf{s}}_{1234n} \hat{\mathbf{s}}_{12n-2n-1n}}
\end{equation*}
\begin{equation*}
    +\frac{1}{\hat{\mathbf{t}}_{4..n-1}} \bigg( \frac{1}{\hat{\mathbf{s}}_{1234n}} + \frac{1}{\hat{\mathbf{s}}_{123n-1n}} \bigg) + \frac{1}{\hat{\mathbf{t}}_{3..n-2}} \bigg( \frac{1}{\hat{\mathbf{s}}_{123n-1n}} + \frac{1}{\hat{\mathbf{s}}_{12n-2n-1n}}\bigg) \bigg]
\end{equation*}
\begin{equation*}
    + \frac{1}{\hat{\mathbf{s}}_{1234n}\hat{\mathbf{t}}_{2..n-3}} \bigg( \frac{1}{\hat{\mathbf{s}}_{12n-2n-1n}} + \frac{1}{\hat{\mathbf{s}}_{1n-3n-2n-1n}} \bigg)
\end{equation*}
\begin{equation*}
    + \frac{1}{\hat{\mathbf{s}}_{1n-3n-2n-1n}\hat{\mathbf{t}}_{4..n-1}} \bigg( \frac{1}{\hat{\mathbf{s}}_{123n-1n}} + \frac{1}{\hat{\mathbf{s}}_{1234n}} \bigg) \Bigg\}
\end{equation*}
\begin{equation*}
    + \frac{1}{\hat{\mathbf{s}}_{1n-3n-2n-1n}\hat{\mathbf{u}}_{4..n}}\bigg[ \frac{1}{\hat{\mathbf{s}}_{12345} \hat{\mathbf{s}}_{123n-1n}}+\frac{1}{\hat{\mathbf{t}}_{5..n}} \bigg( \frac{1}{\hat{\mathbf{s}}_{12345}} + \frac{1}{\hat{\mathbf{s}}_{1234n}} \bigg) 
\end{equation*}
\begin{equation*}
    + \frac{1}{\hat{\mathbf{t}}_{4..n-1}}\bigg( \frac{1}{\hat{\mathbf{s}}_{1234n}} + \frac{1}{\hat{\mathbf{s}}_{123n-1n}} \bigg) \bigg]
\end{equation*}
\begin{equation*}
    + \frac{1}{\hat{\mathbf{s}}_{12345}\hat{\mathbf{u}}_{2..n-2}}\bigg[ \frac{1}{\hat{\mathbf{s}}_{123n-1n} \hat{\mathbf{s}}_{1n-3n-2n-1n}}+\frac{1}{\hat{\mathbf{t}}_{2..n-3}} \bigg( \frac{1}{\hat{\mathbf{s}}_{1n-3n-2n-1n}} + \frac{1}{\hat{\mathbf{s}}_{12n-2n-1n}} \bigg) 
\end{equation*}
\begin{equation*}
    + \frac{1}{\hat{\mathbf{t}}_{3..n-2}}\bigg( \frac{1}{\hat{\mathbf{s}}_{123n-1n}} + \frac{1}{\hat{\mathbf{s}}_{12n-2n-1n}} \bigg) \bigg]
\end{equation*}
\begin{equation}
\label{shiftedk5}
    + \frac{1}{\hat{\mathbf{t}}_{5..n}\hat{\mathbf{t}}_{2..n-3}} \bigg(\frac{1}{\hat{\mathbf{s}}_{12n-2n-1n}} + \frac{1}{\hat{\mathbf{s}}_{1n-3n-2n-1n}} \bigg) \bigg( \frac{1}{\hat{\mathbf{s}}_{12345}} + \frac{1}{\hat{\mathbf{s}}_{1234n}} \bigg).
\end{equation}

\bibliographystyle{JHEP}
\bibliography{references}



\end{document}